\documentclass[reprint,amsmath,amssymb,nofootinbib]{revtex4-2}
\usepackage{graphicx}
\usepackage{xcolor}
\usepackage{amsthm}
\usepackage{amssymb}
\usepackage{stmaryrd}
\usepackage[utf8]{inputenc}

\usepackage{chemfig}
\usepackage{comment}
\usepackage{hyperref}
\usepackage{float}
\usepackage[normalem]{ulem}
\usepackage{wasysym}
\usepackage{siunitx}

\hypersetup{colorlinks = true, linkcolor=blue, citecolor=orange, urlcolor=blue, bookmarksnumbered = true}

\begin{document}

\title{
Local-Limit Disorder Characteristics of Niobium-based\\ Superconducting Radio-Frequency Cavities
}
\author{Anastasiya Lebedeva}
\thanks{These authors contributed equally.}
\author{Mat\'u\v{s} Hladk\'y}
\thanks{These authors contributed equally.}
\author{Marcel Pol\'ak}
\author{Franti\v{s}ek Herman}
\email{herman2@uniba.sk}
\affiliation{Department of Experimental Physics, Comenius University, Mlynská Dolina F2, 842 48 Bratislava, Slovakia}
\begin{abstract}

Nowadays Nb-based superconducting radio-frequency cavities represent fundamental tools used for (Standard Model) particle acceleration, (beyond Standard Model) particle probing, and long-lifetime photon preservation. We study the SRF frequency shift in the vicinity of the critical temperature $T_c$ and the quality factor mainly at low temperatures within the Dynes superconductor model. We scrutinize and use the local limit response to the external electromagnetic field. Our approach allows for a finer analysis of the peculiar behavior of the resonant frequency shift immensely close to $T_c$, observed in recent experiments. In several regimes, we analytically elaborate on the width and depth of the resulting dip. Studying the sign of the slope of the resonant frequency shift at $T_c$ in the moderately clean regime clarifies the role of the pair-breaking and pair-conserving disorder. Next, to find the relevance of our description, we compare and also fit our results with the recent experimental data from the N-doped Nb sample presented by Zarea {\it et al.} [Front. Supercond. Mater. 3, 1 (2023), \url{arXiv:2307.07905}]. Our analysis complies with the experimental findings, especially concerning the dip width. We offer a straightforward, homogeneous-disorder-based interpretation within the moderately clean regime. Comparative analysis for three other cavities with different resonant frequencies reported by Ueki {\it et al.} [Prog. Theor. Exp. Phys. 5, 053I02 (2025), \url{arXiv:2207.14236}] points toward a similar regime. Assuming the same regime at low temperatures, we address details of the high-quality plateaus. Summing all up, this work presents (and studies the limits of) the simple, effective description of the complex problem corresponding to the electromagnetic response in the superconductors, combining homogeneous conventional pairing and two different kinds of disorder scattering.

\end{abstract}

\maketitle

\section*{Introduction}

The response of a superconductor to an external electromagnetic field (resulting in unmeasurable resistivity and the Meissner effect) was crucial from the very beginning of any superconductivity-related research whatsoever. Nowadays, the details of the electromagnetic response of a superconductor are intriguingly important concerning the combination of the low-frequency, microwave external electromagnetic field applied to (most often a conventional) superconductor. This setup is utilized in superconducting radio-frequency (SRF) cavities in particle accelerators \cite{Gurevich17}. Moreover, cavities operating at minimal microwave power and ultralow temperatures $T<\SI{20}{mK}$, with photon lifetimes up to $\SI{2}{s}$, have the potential application in the context of resonators concerning quantum memories or processors~\cite{Romanenko20}. From the experimental point of view, new techniques have been developed to upgrade SRF cavities. It shows, for example, that nitrogen inclusion is an experimentally successful method for enhancing the quality factor up to $Q\sim 6\times10^{10}$ and accelerating gradients up to $\SI{45}{MV/m}$ for superconducting cavities regarding the high-energy particle accelerators \cite{Grassellino13, Grassellino17}. In addition, the nitrogen-doped high-quality factor cavities can serve as very sensitive probes of axion-like, beyond-the-Standard-Model dark matter particle candidates~\cite{Bogorad19, Gao21}.\\

Of course, methods incorporating nitrogen are also interesting from the theoretical point of view, as the microscopic mechanism of doping influences superconducting behavior. In particular, close to the critical superconducting temperature $T_c$, it is interesting to focus on the complex interplay of the electromagnetic response together with the superconducting thermodynamics. So far, several theoretical studies have addressed the method of nitrogen doping. Let us mention a recent theoretical analysis \cite{Ueki22} of the anomalies in the frequency shift measured close to $T_c$, reported in Ref.~\cite{Bafia21}. In the noted manuscript of Ref.~\cite{Ueki22}, the authors address the analysis of the frequency shift and quality factor as functions of temperature, frequency, and disorder. They use the nonequilibrium, microscopic superconducting theory in the weak coupling limit, combined with Slater's method\footnote{Described in detail in the Appendix of Ref.~\cite{Ueki22}.} for solving Maxwell's equations describing the electromagnetic field in a closed metallic cavity \cite{Slater46}. An important effect within this approach is coming from the assumed inhomogeneous distribution of $T_c$, caused by an interplay of gap anisotropy and impurity scattering, leading to weak violation of Anderson's theorem by nonmagnetic disorder \cite{Zarea23b}. In Ref.~\cite{Zarea23a} the authors derive and solve numerically (in general) the complex eigenfrequency equation describing the lowest resonant transverse magnetic mode of a cylindrical cavity. They include penetration and confinement of the electromagnetic field by the normal and superconducting currents in the vicinity of the cavity-vacuum interface. For the current response, they use the Keldysh formulation of the quasiclassical theory of superconductivity \cite{Rainer95}. Within the described approach, disorder effects are considered by the single quasiparticle-impurity mean scattering time parameter. 

In the broader context, the peculiarities in the microwave electromagnetic response are very interesting also outside the Nb-based materials. For example, in Refs.~\cite{Barra05, Barra06}, the authors focus on the role of current redistribution in microwave measurements, giving rise to anomalous effects in the measurements of the surface impedance, penetration depth, and complex conductivity in high-temperature superconducting (HTS) films.\\

In this article, we address peculiarities in a frequency shift $\delta f(T)$ observed in Ref.~\cite{Bafia21} straightforwardly within a simple mean-field-based approach. We exploit the coherent potential approximation (CPA) electron Green's function describing the Dynes superconductor (DS) \cite{Herman16} within Nam's approach of the superconducting response to the electromagnetic field \cite{Nam67a, Nam67b, Herman17b}. 

We take into account the ${\bf k}$-space isotropic superconducting order parameter, together with the effect of the pair-breaking and pair-conserving scattering without any additional inhomogeneity effects. Regarding the anisotropy, its effect can be quantified by the value of factor $\mathcal{A}$ \cite{Ueki22, Zarea23b}. This factor corresponds to the normalized average deviation of the order parameter from the isotropic case on the Fermi surface. Factor $\mathcal{A}$ generally ranges from $\mathcal{A}=0$ for the case of an isotropic s-wave superconductor, up to the extreme of $\mathcal{A}=1$ (case of unconventional superconductors) \cite{Ueki22}. Considering conventional anisotropic superconductors, we find $0<\mathcal{A}<1$, and in the case of Nb, it is reasonable to consider $\mathcal{A}=0.037$ \cite{Ueki22, Zarea23b}. Since the overall deviance from the isotropic case is for most of our considerations under the level of $4\%$, we include its effects within the presence of pair-breaking scattering, slightly reducing the superconducting gap and the critical temperature. The only special case, when our approach strongly requires a highly isotropic superconductor, is considering a very exceptional case with a small pair-breaking but large pair-conserving scattering (ideal dirty regime). In such a case, an order parameter anisotropy would cause a reduction of the $T_c$ not captured correctly by our usually small pair-breaking parameter.

Since we touched on the topic of anisotropy in Nb, it is also reasonable to briefly comment on Nb$_3$Sn. A theory analysis of the Raman-scattering measurements shows $\sim 20\%$ gap anisotropy \cite{Dierker83} in this SRF promising material \cite{PosenPHDThesis}. In such a case, the anisotropy of the order parameter should be taken seriously as it smears the difference between pair-breaking and pair-conserving disorder. However, one can argue by a very rough approximation of the (usually dominating) forward electron scattering acting as a pair-conserving type, meanwhile, the (less-present) backward scattering plays the role of the pair-breaking disorder. It is even possible to average out the results by assuming the anisotropy of the order parameter, which, to some extent, allows us to proceed with our approach.

Since our approach was originally motivated by the Dynes density of states \cite{Dynes78} and therefore naturally reproduces its form \cite{Herman16}; in some cases, it can be viewed as a justification of the phenomenological approaches exploiting the presence of subgap states induced by the Dynes parameter $\Gamma$.

One can also view our approach as an extension of the model, including magnetic (pair-breaking) and nonmagnetic (pair-conserving) impurities throughout the Shiba theory utilized in the context of SRF in Ref.~\cite{Kharitonov12}. However, instead of including just a single coupling constant (equivalent to the Shiba parameter) to the magnetic (or nonmagnetic) impurity, CPA allows for the continuous probability distributions of magnetic and nonmagnetic couplings (for details see Ref.~\cite{Herman16} and references therein). The resulting density of states reproducing the Dynes formula can be viewed as an envelope function of subgap Shiba bands with different coupling strengths of magnetic impurities \cite{HermanPHD, Sindler25}.

The provided article also fits into (and expands) the context of the quality factor analysis from Ref.~\cite{Kubo22} and the detailed discussion of resonant frequency shift and quality presented in Ref.~\cite{Ueki22, Zarea23a}.\\

In our perspective, filling in and discussing the details concerning the DS theory approach, such as understanding the implications of the absence of the coherence peak on the resonant frequency shift behavior, allows searching for the framework usability limits. Such a discussion is crucial if we consider that the DS model already describes the disorder effects from heavily disordered samples being close to the superconductor-insulator transition \cite{Sindler22} to relatively clean materials used in SRF cavities \cite{Herman21}. Note that the mentioned examples stand for the two opposite sides of the disorder spectrum. Focusing on the relevant temperature range, again, we can start with the low-temperature density of states analysis \cite{Dynes78, Szabo16, Herman16}, up to the coherence peak study \cite{Herman21}. Naturally, focusing on the SRF frequency shift measurements, the combination of temperatures extremely close to $T_c$, a relatively low amount of disorder, and microwave electromagnetic wavelengths creates yet another challenge for the framework.\\ 

To be more precise, we try to understand the delicate range in the immediate vicinity of $T_c$ (on the level of $1\permil-1\%$ of its scale), where five effects can, in principle, play a competing role. Superconducting pairing interaction {\it creating the spectroscopic gap $\Delta$ and coherence effects}. Temperature $T$ {\it harshly suppressing superconductivity}. Pair-breaking scattering rate $\Gamma$ {\it introducing the subgap states and moderating the coherence effects}. Pair-conserving scattering rate $\Gamma_s$, {\it which usually dominates the disorder but does not influence the underlying density of states. It also enhances the penetration depth and the coherence effects in conductivity}. Last, not least, the electromagnetic frequency-related energy scale\footnote{For convenience, we use units $\hbar=1$ in derivations. However, when we find it necessary, we use $\hbar$ explicitly, especially when discussing numeric estimates of important quantities.} $\hbar\omega$.\\

Technically, we derive the impedance $Z_s$ properties within the experimentally relevant DS theory \cite{Herman16, Herman17a, Herman17b, Herman18, Herman21, Herman23, Dynes78, Szabo16, Maiwald20, Sindler22}. We discuss and focus mainly on the local limit, representing a scenario when the penetration depth is much larger than the length of the mean free path $\ell$ in normal, or the size of the Cooper pair in the superconducting state, respectively. By studying $Z_s$, we naturally interconnect the optical conductivity $\sigma_s(\omega)$ \cite{Herman17b, Herman21} together with the experimentally accessible surface resistance $R_s = {\mathrm Re}\lbrace Z_s\rbrace$ (and, therefore, also quality factor $Q_s \propto 1/R_s$) and the resonant frequency shift $\delta f(T)\propto{\mathrm Im}\lbrace Z_s \rbrace$. Among other topics, we also discuss recent experimental results \cite{Bafia21} on resonant frequency shift of superconducting cavities from the point of view of our approach.\\

To present our findings clearly and systematically, we organize the material in the following way. In Sec.~\ref{sec:Imp&Cond}, we review the role of the superconducting optical response in the surface impedance. Next, in Sec.~\ref{sec:FreqShift} we get to the related frequency shift, and in Sec.~\ref{sec:Q} - to the quality factor analysis. The final Sec.~\ref{sec:Concl&Discuss} is dedicated to our conclusions and their discussion. Additional material can be found in the Appendices~\ref{app:Dip_2FM}-\ref{Appendix:App_Nb3Sn}.

\section{Surface impedance And microwave conductivity}\label{sec:Imp&Cond}
\subsection{Surface Impedance $Z(\omega)$}\label{subsec:Z}

We study the interplay of the electromagnetic field on the surface of a normal metal or a superconductor. Therefore, we focus on the surface impedance $Z(\omega)$, expressing the semi-infinite medium interface solution of the Maxwell equations, neglecting the effect of the displacement current \cite{Nam67b, Herman21} and assuming (strictly speaking\footnote{The diffuse reflection leads to similar result \cite{Callaway74}.}) specular reflection of the electrons at the interface in the following form \cite{Nam67b, Callaway74}:
\begin{equation}\label{eq:Impedance_def}
    Z(\omega) = \frac{i\omega\mu_0}{\pi}\int_{-\infty}^{\infty}\frac{dq}{q^2 + i \omega\mu_0\big(\sigma'(q,\omega)-i\sigma''(q,\omega)\big)}.
\end{equation}
Medium specifics enter the response throughout the transverse electron conductivity $\sigma(q,\omega)$ being (in general) a function of the wavevector $q$ and frequency $\omega$. Constant $\mu_0$ is the vacuum permeability. Moreover, the impedance can be divided into surface resistance $R(\omega)$ and surface reactance $X(\omega)$ \cite{Slater46, Ueki22}, both experimentally accessible quantities \cite{Bafia21}.

For simplicity, we focus on the resulting form of Eq.~\eqref{eq:Impedance_def} in two limit cases. In the local, London limit $(q\rightarrow 0)$ $\sigma(q,\omega)\approx \sigma_{n,s}(\omega)$ in the normal or superconducting state, and Eq.~\eqref{eq:Impedance_def} can be calculated straightforwardly in the complex plane \cite{Nam67b, Herman21}. In the anomalous, Pippard limit $($assuming $q\rightarrow \infty$ and $\omega\tau\rightarrow 0$, where $\tau$ is the scattering time$)$, $\sigma(q,\omega)\propto \sigma_{n,s}(\omega)/q$ \cite{Nam67b} and the resulting integral can be also found analytically \cite{Callaway74}. In Sec.~\ref{sec:Temp_Range}, we briefly discuss the relevance of both: local and anomalous normal state responses. However, a detailed analysis of the frequency shift utilizing DS theory will be provided in the more relevant, local limit.

\begin{figure}[h!]
\includegraphics[width = 6.1 cm]{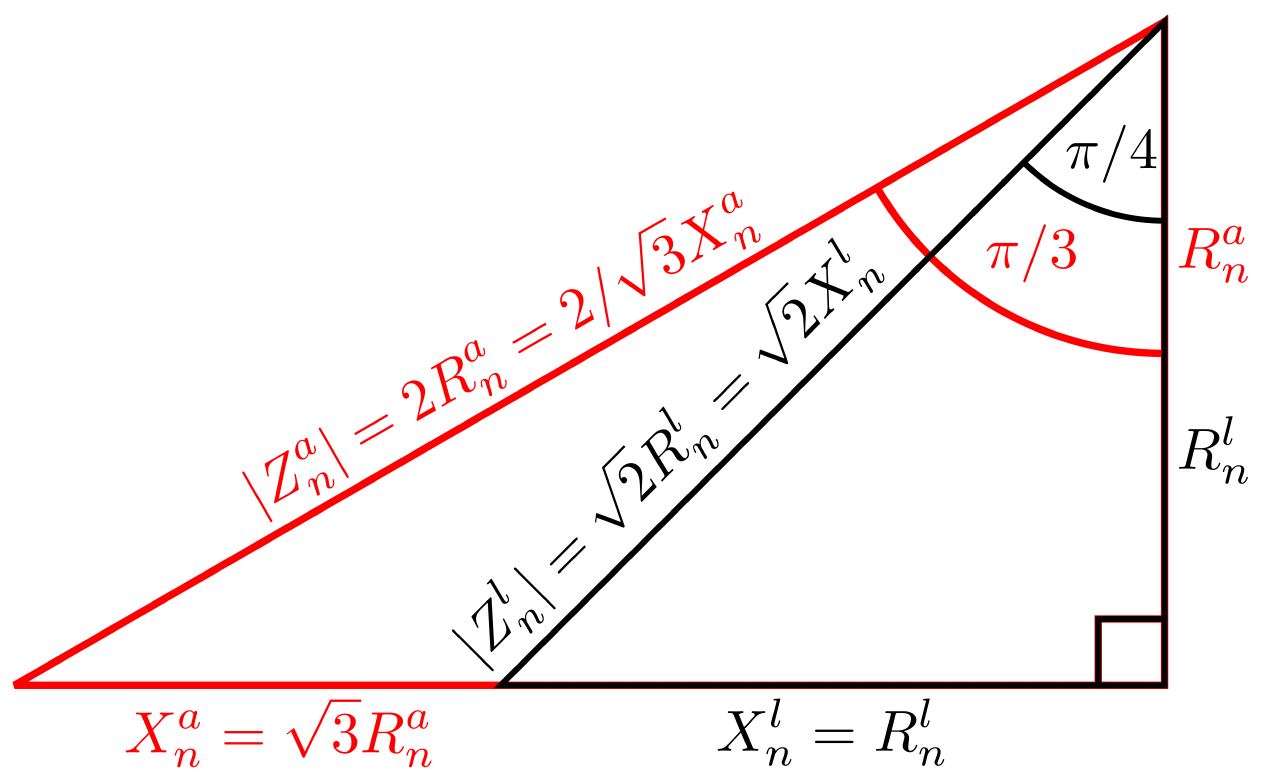}
\caption{Right-angle triangle diagram of the surface resistance and reactance in the normal state. The black triangle corresponds to the local limit; meanwhile, the red triangle corresponds to the anomalous limit. Angle values in the upper corner correspond to $\alpha_n^{l}$, respectively $\alpha_n^{a}$ for $\omega\tau\rightarrow 0$. Just for illustration purposes, we consider $R_n^l=R_n^a$ .}
\label{fig:ZnlZna}
\end{figure}

For our purposes, in both limit cases, it seems to be convenient to express the superconducting impedance $Z_s(\omega)$ throughout the ratio of superconducting $\sigma_s(\omega)$ and normal state $\sigma_n(\omega)$ conductivities in terms of amplitudes and phases as \cite{Gurevich17, Gurevich23, Nam67b, Mattis58, Parks97}
\begin{align}\label{eq:Impedance}
    Z_s &\equiv R_s + i X_s = Z_n \frac{R_s + i X_s}{R_n + i X_n} = Z_n\left(\frac{\sigma_s'-i\sigma_s''}{\sigma_n'-i\sigma_n''}\right)^k, \nonumber \\
        &= |Z_n| e^{i\alpha_n} \left(\frac{|\sigma_s|}{|\sigma_n|}\right)^k e^{-i k \delta \varphi},
\end{align}
where $|Z_n|~=~\sqrt{R_n^2 + X_n^2}$, $\alpha_n~=~\arctan{\big(X_n/R_n\big)}$, $\delta\varphi=\varphi_s-\varphi_n$ is the conductivity phase difference and exponent $k = -1/2$ ($k=-1/3$) in local (anomalous) limit \cite{Nam67b}. For further purposes, it is straightforward to identify superconducting surface resistance $R_s(\omega)$ and reactance $X_s(\omega)$ from Eq.~\eqref{eq:Impedance} as
\begin{align}
    \frac{R_s}{|Z_n|} &= \left(\frac{|\sigma_s|}{|\sigma_n|}\right)^k \cos(\alpha_n- k \delta \varphi),\label{eq:Rs} \\
    \frac{X_s}{|Z_n|} &= \left(\frac{|\sigma_s|}{|\sigma_n|}\right)^k \sin(\alpha_n- k \delta \varphi), \label{eq:Xs}
\end{align}
with expected normal state limit form of
\begin{equation*}
R_n=|Z_n|\cos{(\alpha_n)}\quad\textrm{and}\quad X_n=|Z_n|\sin{(\alpha_n)}.
\end{equation*} 
Moreover, in the limit $\omega\tau\rightarrow 0 \quad \sigma_n''\rightarrow 0$, and the relation between local $(X_n^{l}, R_n^{l}, Z_n^{l},\alpha_n^{l})$, respectively anomalous $(X_n^{a}, R_n^{a}, Z_n^{a},\alpha_n^{a})$ quantities can be nicely presented in the form of a right-angle triangle diagram as shown in Fig.~\ref{fig:ZnlZna}.

In the following, we focus on Slater's solution of Maxwell's equations with the effect of damping, caused by the resistive losses on the SRF hollow cavity walls  \cite{Slater46, Ueki22}. We neglect the effects of the waveguide outputs and (naturally) also current sources in the vacuum region \cite{Ueki22, Trunin97, Shevchun06}. The relation between the quality factor $Q$, resonant angular frequency shift $\Delta \omega_m$ of the {\it m}th mode in an ideal resonant cavity, and the surface impedance $Z$ can be expressed as
\begin{equation}\label{eq:Q&AngShift}
    \frac{1}{Q} - 2i\frac{\Delta \omega_m}{\omega_m} = \frac{Z}{G},\quad G = \frac{\omega_m Z_0}{c}\frac{\int {\bf H}\cdot{\bf H}_m dV}{\int {\bf H}\cdot{\bf H}_m dS}.
\end{equation}
The value of $G$ is sensitive to the realized geometry of the cavity through the volume. It is related to the (conducting) surface integrals of the magnetic field strength ${\bf H}$ being expanded to {\it m}th solenoidal mode ${\bf H}_m$. $Z_0=\sqrt{\mu_0/\varepsilon_0}\approx\SI{376.7}{\Omega}$ ($c=1/\sqrt{\mu_0\varepsilon_0}\approx2.998\times10^8\SI{}{m/s}$) is the so-called vacuum impedance (speed of light). For the standard SRF Tesla cavities $G\approx 270 \, \Omega$ \cite{Aune00, Bafia20}.\\ 

\subsection{Microwave conductivity $\sigma(\omega)$}

{\it (Normal state)} In the previous Sec.~\ref{subsec:Z}, we identified the role of conductivity $\sigma_{n,s}(\omega)$ in the superconducting surface resistance $R_s(\omega)$ and reactance $X_s(\omega)$, formulated by Eq.~\eqref{eq:Rs} and Eq.~\eqref{eq:Xs}. Since our analysis of the frequency shift focuses on the temperatures close to $T_c$, it is natural to relate the considered effects to the normal state, assumed to be in the local limit (this assumption will be discussed in the Sec.~\ref{sec:Temp_Range}). Therefore, the $\omega$ dependence of the a.~c. normal state conductivity $\sigma_n(\omega)$ will be taken in the form of the Drude formula as
\begin{equation}\label{eq:Drude}
    \sigma_n(\omega)/\sigma_0 = (1-i\omega\tau)^{-1}=\left(1+\left(\omega\tau\right)^2\right)^{-1/2} e^{i\varphi_n}.
\end{equation}
The usual relation between the scattering time $\tau$ and normal state scattering constant $\Gamma_n$: $2\tau=\hbar/\Gamma_n$ holds true. In addition, $\sigma_0 = ne^2\tau/m$, where $n$ is particle density, $e$ is the electric charge, and $m$ is the electron mass. In Eq.~\eqref{eq:Drude}, we also expressed $\sigma_n(\omega)$ using a bit more convenient module and phase form, defining $\varphi_n = \arctan\left(\omega\tau\right)$.\\

Using Eq.~\eqref{eq:Drude}, we consider the normal state impedance $Z_n$ in the local limit $(k=-1/2)$ \cite{Gurevich17} as
    \begin{align}
        Z_n &= \sqrt{\frac{i\mu_0 \omega}{\sigma'_n - i\sigma''_n}},\nonumber\\
        &= R_n\left(
        \sqrt{\sqrt{1+\left(\omega\tau\right)^2}-\omega\tau} + i\sqrt{\sqrt{1+\left(\omega\tau\right)^2}+\omega\tau}
        \right),\nonumber\\
        &= \sqrt{2} R_n\sqrt[4]{1+(\omega\tau)^2}e^{i\alpha_n}\label{eq:Zn},
    \end{align}
    where $\alpha_n = \pi/4 + \arctan\left(\omega\tau\right)/2$,
    and in the limit of $\omega\tau \rightarrow 0$, we get the textbook result of $R_n~=~X_n~=~\sqrt{\mu_0\omega/(2\sigma_0)}$ \cite{Callaway74}.\\

{\it (Superconducting state - Dynes superconductor)} Our formulation of the superconducting state is motivated by the microscopic interpretation of the Dynes formula for the tunneling density of states \cite{Dynes78, Herman16}. It is constructed within the convenient language of Nambu-Gorkov superconducting Green's functions within the CPA approach \cite{Soven67, Velicky68, Velicky69, Janis21}. The solution of the CPA equations results in the convenient form of the gap function $\Delta(\omega)$ and wavefunction renormalization $\mathrm{Z} (\omega)$ \cite{Herman17b, Herman21}

\begin{equation*}
    \Delta(\omega) = \overline{\Delta} \bigg/ \left(1 + \frac{i\Gamma}{\omega}\right),\, \mathrm{Z}(\omega) = \left(1 + \frac{i\Gamma}{\omega}\right)\left(1 + \frac{i\Gamma_s}{\Omega(\omega)}\right),
\end{equation*}
where\footnote{We choose the square root branch so that the signs of $\mathrm{Re}\{\Omega(\omega)\}$ and $\omega$ are the same.} $\Omega(\omega) = \sqrt{\left(\omega + i\Gamma\right)^2 - \overline{\Delta}^2}$ and $\overline{\Delta}=\overline{\Delta}(T)$. This formulation results in the Dynes tunneling density of states, fulfills the Anderson theorem \cite{Anderson59} and results in a natural normal state limit $\Delta(\omega)=0$, $\mathrm{Z}_n(\omega) = (1+i\Gamma_n/\omega)$, where $\Gamma_n = \Gamma_s + \Gamma$ unifies the effect of the disorder, as expected.\\

Notation note: Within the following sections, we often use experimentally convenient units of the determined gap of the disordered system [known, for example, from scanning tunneling microscopy (STM) measurements at $T=\SI{0}{K}$] $\Delta_0~=~\overline{\Delta}(0)~=~\Delta_{00}\sqrt{1-2\Gamma/\Delta_{00}}$ \cite{Herman16}, where $\Delta_{00}$ is the ideal, Bardeen-Cooper-Schrieffer (BCS)-like superconducting gap. The thermodynamic properties of the DS theory \cite{Herman18, Herman23} show that the well-known BCS ratio $\Delta_{00}/T_{c,0}~\approx~1.764$ modifies in the limit of the critical $\Gamma\rightarrow\Delta_{00}/2$ to the Dynes critical ratio $\Delta_{0}/T_c~=~\sqrt{2/3}\pi \approx~2.565$ \cite{Lebedeva24}. We also work with the nondimensional scattering rates $\gamma \equiv \Gamma/\Delta_0$ and $\gamma_s \equiv \Gamma_s/\Delta_0$. In addition, as expected, our general results simplify in various regimes of low and high scattering rates. Due to the difference in their nature, we will note them as ideal ($\gamma \ll 1$) or bad ($\gamma \gg 1$) and clean ($\gamma_s \ll 1$) or dirty ($\gamma_s \gg 1$), similarly to Ref.~\cite{Kubo22}.\\

When discussing the optical response of superconductors, it is
customary to follow Nam \cite{Nam67a} and introduce three
complex functions: density of states $n(\omega)$ and density of pairs $p(\omega)$, defined by
\begin{eqnarray}
n(\omega)&=&n_1(\omega)+in_2(\omega)
=\frac{\omega}{\sqrt{\omega^2-\Delta^2(\omega)}}=\frac{\omega+i\Gamma}{\Omega(\omega)},\label{eq:n(w)}\\
p(\omega)&=&p_1(\omega)+ip_2(\omega)
=\frac{\Delta(\omega)}{\sqrt{\omega^2-\Delta^2(\omega)}}=\frac{\overline{\Delta}}{\Omega(\omega)},\label{eq:p(w)}
\end{eqnarray}
and energy function $\epsilon(\omega)$
\begin{align}
\epsilon(\omega)=\epsilon_1(\omega)+i\epsilon_2(\omega)
&=\mathrm{Z}(\omega)\sqrt{\omega^2-\Delta^2(\omega)}\nonumber\\
&=\Omega(\omega) + i \Gamma_s.\label{eq:e(w)}
\end{align}

The functions $n(\omega)$ and $p(\omega)$ are obviously
not independent, and they satisfy the constraint
$n^2(\omega)~-~p^2(\omega)~=~1$. Note that the
functions $n_1(\omega)$, $p_2(\omega)$, and $\epsilon_2(\omega)$ are
even, whereas the functions $n_2(\omega)$, $p_1(\omega)$ and $\epsilon_1(\omega)$ are odd in $\omega$. Next, it is well known \cite{Carbotte05, Herman17b, Herman21} that the finite-frequency
part of the local optical conductivity
$\sigma_s(\omega)=\sigma_s(0,\omega)$ of a superconductor is given
by
\begin{eqnarray}
\sigma_s(\omega)=\frac{iD_0}{\omega}\int_{-\infty}^{\infty}d\nu
\tanh\left(\frac{\nu}{2T}\right) H(\nu+\omega,\nu).
\label{eq:regular}
\end{eqnarray}
Here $D_0=ne^2/m$ is the normal-state Drude weight and
\begin{align}
H_1 \big(x, y\big) &=
\frac{1 + n(x)n^*(y) + p(x)p^*(y)}
{2\left[\epsilon^*(y) - \epsilon(x)\right]}, \nonumber\\
H_2 \big(x, y\big) &= \frac{1 - n(x)n(y) - p(x)p(y)}
{2\left[\epsilon(y)+\epsilon(x)\right]}, \nonumber\\
H \big(x, y\big) &= H_1 \big(x, y\big) + H_2 \big(x, y\big),
\label{eq:function_h}
\end{align}
are auxiliary complex functions exploiting Eq.~(\ref{eq:n(w)}-\ref{eq:e(w)}). Even though the formulation using Eq.~(\ref{eq:regular}, \ref{eq:function_h}) is elegant, we can find a form of $\sigma_s(\omega)$ more adjusted to numerical considerations because of the unpleasant asymptotic behavior of $\tanh(\nu)$ and $H_{1}\big(\nu + \omega, \nu\big)$ for $\nu\rightarrow\pm\infty$. By assuming symmetries
\begin{align*}
    H_{1}^*\left(x, y\right) &= -H_{1}\left(-x,-y\right) = - H_{1}\left(y,x\right), \\
     H_{2}^*\left(x, y\right) &= -H_{2}\left(-x,-y\right)= H_{2}^*\left(y,x\right).
\end{align*}
Eq.~\eqref{eq:regular} can be written as
\begin{multline}\label{eq:Sigma_Num}
    \sigma_s(\omega)=\frac{iD_0}{\omega}\int_{-\infty}^{\infty}d\nu \Big[ \big(f(\nu+\omega)-f(\nu)\big)H_1(\nu+\omega,\nu)\\
     -f(\nu)\big(H_2(\nu+\omega,\nu)+H_2^*(\nu,\nu-\omega)\big)\Big],
\end{multline}
where we utilized the Fermi-Dirac distribution $f(\nu)~=~1/(1+e^{\nu/\left(k_BT\right)})$.

\subsection{Local (London) and Anomalous (Pippard) Relevant Temperature Ranges}\label{sec:Temp_Range}

{\it (Normal state)} To distinguish local and anomalous regimes in the normal state, usually, the penetration depth of the field $\lambda_{f}$ is compared with the mean free path of the electrons $\ell$ \cite{Gurevich17}. In the case of $\lambda_{f}\gg \ell$, the electron scattering on disorder bounds the electron interactions with the field to the local limit. On the other hand, in the case of $\lambda_{f}\ll \ell$, only a fraction of electrons, traveling parallel and near the surface, will respond to the electromagnetic field in the so-called anomalous response regime \cite{Callaway74}. As an estimate (exact in the local limit) of $\lambda_f$, one can look at (inverse of) the characteristic value of the wavevector in Eq.~\eqref{eq:Impedance_def} defined as $|q_0|=\sqrt{\omega\mu_0|\sigma(\omega)|}$. If, for simplicity, we assume Drude-like form for $\sigma(q,\omega)\approx \sigma_n(\omega)$ defined by Eq.~\eqref{eq:Drude}, we are left with \cite{Herman21}
\begin{equation}\label{eq:q0l}
    |q_0| \ell =  \frac{\ell}{\lambda_{L0}}\left(\frac{(\omega\tau)^2}{1+(\omega\tau)^2}\right)^{1/4}\approx \frac{\sqrt{\omega\tau}\ell}{\lambda_{L0}}=\frac{\sqrt{2}\ell}{\delta},
\end{equation}
where $\lambda_{L0} = \sqrt{m/(\mu_0 n e^2)}$ is shown to be identical to the London penetration depth of the superconductor in the clean limit at $T = \SI{0}{K}$. For explanatory purposes, we also utilize the skin effect penetration depth $\delta=\lambda_{L0}\sqrt{2/\omega\tau}$ throughout the article. We also assume the regime $\omega\tau~\ll~1$ in the second step of Eq.~\eqref{eq:q0l}. Note also that by using $\lambda_{L0}$, normal state resistance and reactance in Eq.~\eqref{eq:Zn} can be written as 
\begin{equation}\label{eq:Rn_or_X_n}
R_n~=~X_n~=~\omega\mu_0\delta/2=~\mu_0\lambda_{L0}\sqrt{\omega/(2\tau)}.
\end{equation}

To quantify $|q_0| \ell$, we use the following numbers: resonant frequency $f~=~\SI{1.3}{GHz}$ \cite{Bafia21}, pure Nb Fermi velocity $v_F~\approx~\SI{0.3}{Mm/s}$ \cite{Pronin98, Herman21} and order of magnitude estimate for $\ell\sim \SI{100}{nm}$ \cite{Bafia20}. We are left with the scattering time estimate $\tau=\ell/v_F\approx\SI{0.3}{ps}$, and $\omega\tau \approx 2.7\times 10^{-3}$. Next, assuming $T~=~\SI{0}{K}$ and penetration depth of the pure niobium $\lambda_{L0} = \SI{39}{nm}$ \cite{Bafia21, Maxfield65}, we are left with $|q_0|\ell \approx 0.13$, roughly satisfying the local limit condition $|q_0|\ell \ll 1$ \cite{Callaway74}. However, considering $\ell~\sim~\SI{700}{nm}$, we are left with $\tau=\SI{2.3}{ps}$, $\omega\tau\approx1.9\times10^{-2}$ and $|q_0|\ell \approx 2.48$, which starts to point toward the anomalous behavior $|q_0|\ell \gg 1$~\cite{Callaway74}, however, it is not fully there. 

Since we are already working with relevant numbers, we can also estimate the relevant range of values for the dominant pair-conserving scattering rate $\Gamma_s$, since $\tau=\hbar/(2\Gamma_n)$. Neglecting the less-abundant (however important) pair-breaking scattering leads to $\Gamma_n~\approx~\Gamma_s~=~\hbar v_F/(2 \ell)~\approx~\SI{1}{meV}$ in the case of the smaller considered $\ell$, respectively $\Gamma_s\approx\SI{0.14}{meV}$ in the case of larger $\ell$. Since the $\SI{}{meV}$ scale is the scale of the superconducting gap $\Delta_0$, it is reasonable to expect $\Gamma_s\in(0.1,1)\Delta_0$.

{\it (Superconducting state)} In the superconducting regime, we can assume the relevant scale distinguishing the local and anomalous response to be $|q_0|\xi$, where $\xi$ is the size of the Cooper pair. If we assume that the pairing interaction localizes the electromagnetic response, $|q_0| \xi \ll 1$ corresponds to the local limit. Starting at $T=\SI{0}{K}$ and realizing dominant part of $\sigma_s''$, we are left with $1/ |q_0|=\lambda_L(0)=\lambda_{L0}(1+\xi_0/\ell)^{1/2}$ and $\xi \approx \xi_0/(1+\xi_0/\ell)$ \cite{Herman21, Herman23} for which 
\begin{equation}\label{eq:sc_cond_loc_anomal}
    |q_0| \xi \approx\frac{\xi_0}{\lambda_{L0}}\left(1+\frac{\xi_0}{\ell}\right)^{-3/2}.
\end{equation}
Next, assuming the gap of the clean sample  $\Delta_{00}~\sim \SI{2}{meV}$ at $T=\SI{0}{K}$, the size of the Cooper pair results in $\xi_0~=~\hbar v_F/(\sqrt{8}\Delta_{00})~\approx~\SI{35}{nm}$. Note that this scale is very similar to the scale of $\lambda_{L0}$ and that the ratio $\xi_0/\ell=\Gamma_s/(\sqrt{2}\Delta_{00})$ gives another way to estimate the value of pair-conserving scattering rate. 

However, connecting this argument with the idea from the previous paragraph reveals that this value does not significantly affect the result of Eq.~\eqref{eq:sc_cond_loc_anomal} being $|q_0|\xi~\approx~ 0.57$, (respectively. $0.83$) in the case of smaller (respectively, larger) $\ell$. Both values point toward the local limit behavior (being fully justified in the regime $|q_0| \xi \ll 1$ to be precise).

Next, let us use the fact that the penetration depth $\lambda_L(T)~\propto~1/\sqrt{1-(T/T_c)^4}$ \cite{Tinkham} increases, whereas the size of the Cooper pair changes only very slightly with temperature \cite{Herman23}. Decreasing $|q_0|\xi$ supports the local limit regime up to the normal state if the latter is local. However, the anomalous regime in the normal state does not guarantee the local regime for temperatures up to $T_c$. In our considerations, we can end up in both cases, depending on the value of $\ell$. Hence, it is interesting to look at the temperature scale $T_{s/a}$, where the anomalous regime starts to play a role in the superconducting state. For that, we use a rough estimate coming from the comparison of the superconducting-local surface impedance, dominated by the imaginary part of the conductivity $Z_s~\sim~i\omega\mu_0\lambda_L(T)$, and the purely anomalous normal state impedance $Z_a\sim(1+i\sqrt{3})\omega\mu_0 L_{\omega}$ \cite{Callaway74}, where $L_{\omega}^3~=~\lambda_{L0}^2v_F/\omega$. The very rough criteria estimating $T_{s/a}$ can be formulated \cite{Herman21} by looking at the case $|Z_a|<|Z_s|$. This inequality signifies the failure of the picture considering only limiting regimes and the need for a more general description. It results in
\begin{equation}\label{eq:Crit}
\left(\frac{\lambda_{L0}}{\lambda_{L}(T)}\right)^2=\frac{n_s(T)}{n}<\frac{1}{4}\left(\frac{\omega \lambda_{L0}}{v_F}\right)^{2/3}.
\end{equation}
To further identify $T_{s/a}$, we focus on the temperature dependence of the ratio of the superconducting to normal state particle densities close to $T_c$ as $n_s(T)/n~\approx~A(1~-~T/T_c)$. The coefficient $A\approx2$ weakly depends on the values of the scattering rates $\Gamma$ and $\Gamma_s$~\cite{Herman21}. Combining it with Eq.~\eqref{eq:Crit} leads to
\begin{equation}\label{eq:T_sa}
    \frac{T_{s/a}}{T_c}\equiv 1-\frac{1}{4A}\left(\frac{\omega \lambda_{L0}}{v_F}\right)^{2/3}\leq\frac{T}{T_c}\leq1.
\end{equation}
Considering resonant frequency $f~=~\SI{1.3}{GHz}$ and other already introduced parameters, we obtain $T_{s/a}~\approx~0.9987T_c$. Note that $T_{s/a}$ is not significantly below the relevant temperature ranges corresponding to the peculiar frequency shift behaviors described in Sec.~\ref{sec:FreqShift}, even when we consider that the normal state is strongly anomalous. Since $T_{s/a}$ is comparable or even closer to $T_c$ than the discussed temperature scale, it is natural to assume that local limit behavior still plays an important role in our considerations. Therefore, for simplicity, we focus only on the local limit behavior and we neglect the relevance of the local/anomalous transition. We are motivated to do so by the values from the normal state considerations $|q_0|\ell\approx0.13$ (respectively $|q_0|\ell\approx2.48$). These values either point toward the local normal state or, considering the case of the cleaner sample, we are still not fully in the anomalous limit. Despite working purely within the local limit, we do not want to hold the role of the mixed regime out, and it deserves attention in the future.

From Eq.~\eqref{eq:T_sa} we note that the ratio $T_{s/a}/T_c$ depends on the scattering rates through the value of the coefficient $A$. Yet, it seems that it does not depend on the value of $\ell$ directly, as one could expect. However, Eq.~\eqref{eq:T_sa} can be rewritten with the help of Eq.~\eqref{eq:q0l} as
\begin{equation*}
    \frac{T_{s/a}}{T_c} = 1-\frac{\omega\tau}{4A(|q_0|\ell)^{2/3}}.
\end{equation*}
Here $T_{s/a}$ correctly scales with increasing $A$, meaning we have more (locally behaving) superconducting particles in the system. In addition, $T_{s/a}\rightarrow T_c$ as $\omega\tau\rightarrow0$, since the scattering time scale is much shorter than the period of the alternating electromagnetic field supporting the local limit behavior, as already seen in Eq.~\eqref{eq:q0l}. 

\section{Frequency shift}\label{sec:FreqShift}

Among other measurables, the difference between the superconducting and normal state is represented by the resonant frequency shift of the superconducting cavity, defined as $\delta f \equiv \Delta f_{m,s}-\Delta f_{m,n}$ (realizing $\omega=2\pi f$). Assuming resonant cavity with small damping $\Delta f_m\ll f_m$, in the limit of large quality $1\ll Q$, $\delta f(T) $ can be expressed using the difference between the imaginary parts of Eq.~\eqref{eq:Q&AngShift} and Eq.~\eqref{eq:Impedance} in the superconducting and normal state as \cite{Trunin97, Shevchun06, Slater46, Ueki22}
\begin{equation}\label{eq:ImpDif}
    X_s(T) - X_n= -2G \frac{\delta f(T)}{f}.
\end{equation}
Defining the convenient frequency unit scale $\tilde{f}~\equiv~f X_n/(2G)$ and combining Eq.~\eqref{eq:ImpDif} together with Eq.~\eqref{eq:Xs} in the local limit
\begin{align}
    \frac{\delta f(T)}{\tilde{f}} & = 1 - \frac{X_s(T)}{X_n},\nonumber\\
                    & = 1 -\sqrt{\frac{|\sigma_n|}{|\sigma_s(T)|}}\frac{\sin\left(\alpha_n+\delta\varphi(T)/2\right)}{\sin\left(\alpha_n\right)},\label{eq:ResFreqShift}\\
                    &\approx 1 - \sqrt{\frac{2\sigma_0}{|\sigma_s(T)|}}\sin\left(\pi/4+\varphi_s(T)/2\right)\left(1 - \omega\tau/2\right)\label{eq:delta_f_loc}.
\end{align}
In the third line, we linearized our result considering the most often relevant limit of $\omega\tau\ll1$.
\begin{figure}[h!]
    \includegraphics[width = 0.32\textwidth]{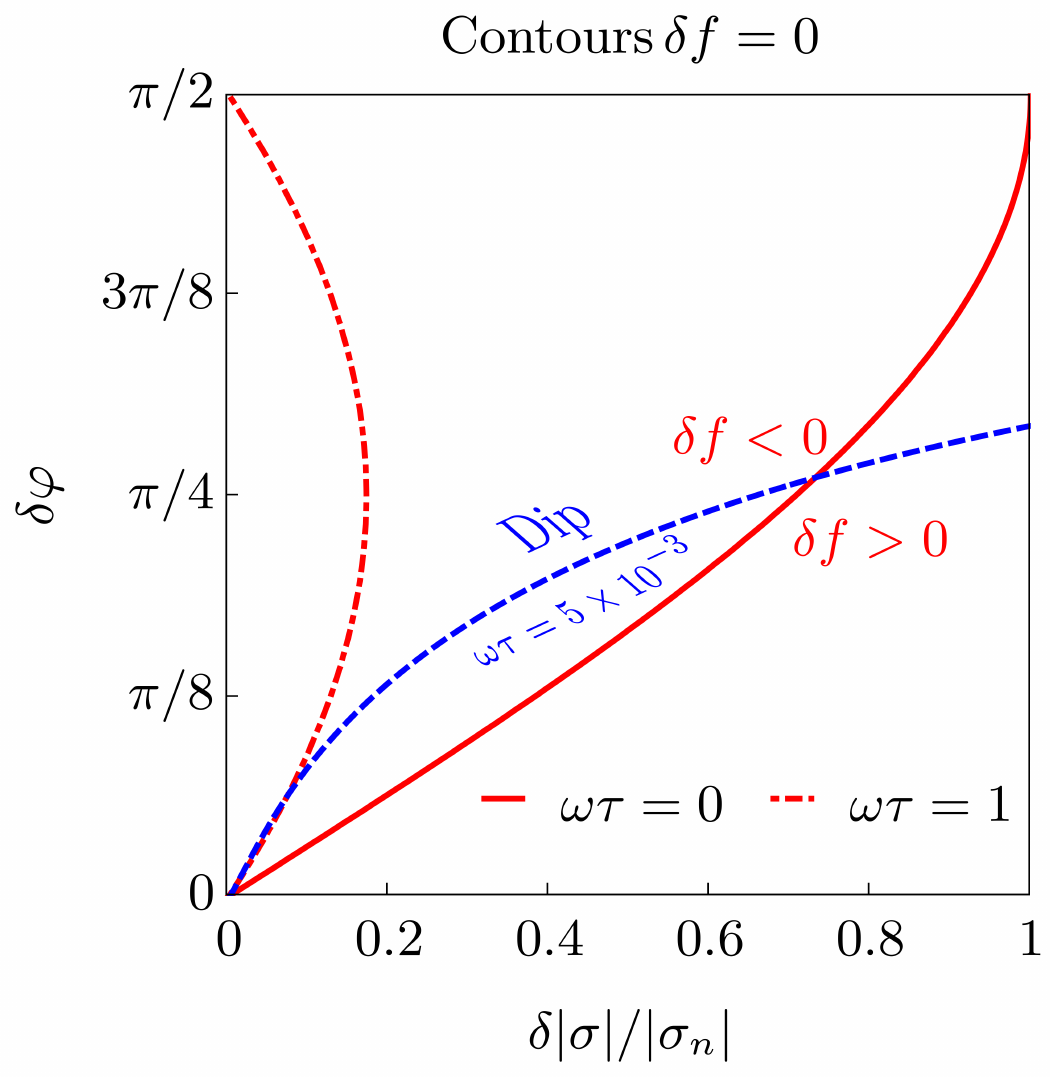}
    \caption{Contour plot $\delta f (T) = 0$ (red curves). The example shown by the dashed blue curve corresponds to the {\it Dip} scenario.}
    \label{fig:Contour}
\end{figure}

In Fig.~\ref{fig:Contour}, we schematically show the behavior of Eq.~\eqref{eq:ResFreqShift} in the form of the contour plot in units (convenient close to $T_c$) of the difference of the superconducting and normal state conductivities $\delta |\sigma|\equiv |\sigma_s| - |\sigma_n|$ and the phase difference $\delta \varphi$. The red contours $\delta f=0$ represent two different choices of $\omega\tau$ for the local limit. The region on the left from the contour for the specific case corresponds to $\delta f<0$, and the region to the right corresponds to $\delta f>0$. Different theories, or experimental data, draw lines (as shown by the {\it Dip} example) parametrized by the increasing temperature going from the superconducting state (top right region of the figure), to the normal state located at point $(0,0)$. The resonant frequency shift will switch signs, as the curve crosses the contour. The linearized form of Eq.~\eqref{eq:ResFreqShift} in the language of small $\delta|\sigma(T)|/|\sigma_n|$ and $\delta \varphi(T)$ reads
\begin{equation*}
    \frac{\delta f(T \rightarrow T_c)}{\tilde{f}} 
    \approx \frac{1}{2}\left(\frac{\delta |\sigma(T)|}{|\sigma_n|}-\cot(\alpha_n)\delta\varphi(T)\right),
\end{equation*}
explaining the rising slope of contours close to point $(0,0)$ with $\omega\tau$ in Fig.~\ref{fig:Contour}.

Next, considering case of $\omega\tau\ll1$, where $\sigma'_n \gg \sigma''_n$ and the temperatures close to $T_c$, where $\sigma'_s~\gg~\sigma''_s$, we are left with
\begin{align*}
    \frac{\delta f(T \lessapprox T_c)}{\tilde{f}} 
    &\approx \frac{1}{2}\left(\frac{\delta \sigma'(T)}{\sigma_0}-\delta\varphi(T)\right),\\
    \end{align*}
where $\delta\sigma'(T)=\sigma'_s(T)-\sigma'_n$. Let us also assume that $\delta\sigma'(T)/\sigma_0\geq0$ due to the presence of the coherence peak. Under these assumptions, the analysis leads to the {\it Dip} feature ($\delta f (T\lessapprox T_c)<0$) close to $T_c$, if $\delta\sigma'(T)/\sigma_0$ is smaller than the phase change $\delta \varphi(T)$. As a follow-up, let us focus on the connection between the absence of the coherence peak and the presence of the {\it Dip} structure in $\delta f(T)$ right under $T_c$. Note that since $\delta\varphi(T) \geq 0$ for $T \leq T_c$, the absenting coherence peak in $\delta\sigma'(T)/\sigma_0 \leq 0$ (suppressed by the influence of pair-breaking effects \cite{Herman21}) leads to $\delta f(T) \leq 0$.

To complete our discussion and determine the relevant scale of $\delta f(T)$ at low temperatures, we can analyze Eq.~\eqref{eq:delta_f_loc}, assuming $\omega\tau\rightarrow 0$ for $T\ll T_c$, where usually $\sigma'_s/\sigma''_s\ll1$, resulting in
\begin{equation*}
    \frac{\delta f(T)}{\tilde{f}} \approx 1-\sqrt{\frac{2\sigma_0}{\sigma_s''(T)}}\left(1-\frac{3}{8}\left(\frac{\sigma_s'(T)}{\sigma_s''(T)}\right)^2\right).
\end{equation*}
Assuming the lowest interesting order at \SI{0}{K} temperature, we find $\delta f(0)/\tilde{f}\approx1-2\lambda_L(0)/\delta$.

\subsection{Imprints of various regimes in local limit}\label{subsec:imprints}

\begin{figure}[h!]
    \includegraphics[width = 0.36\textwidth]{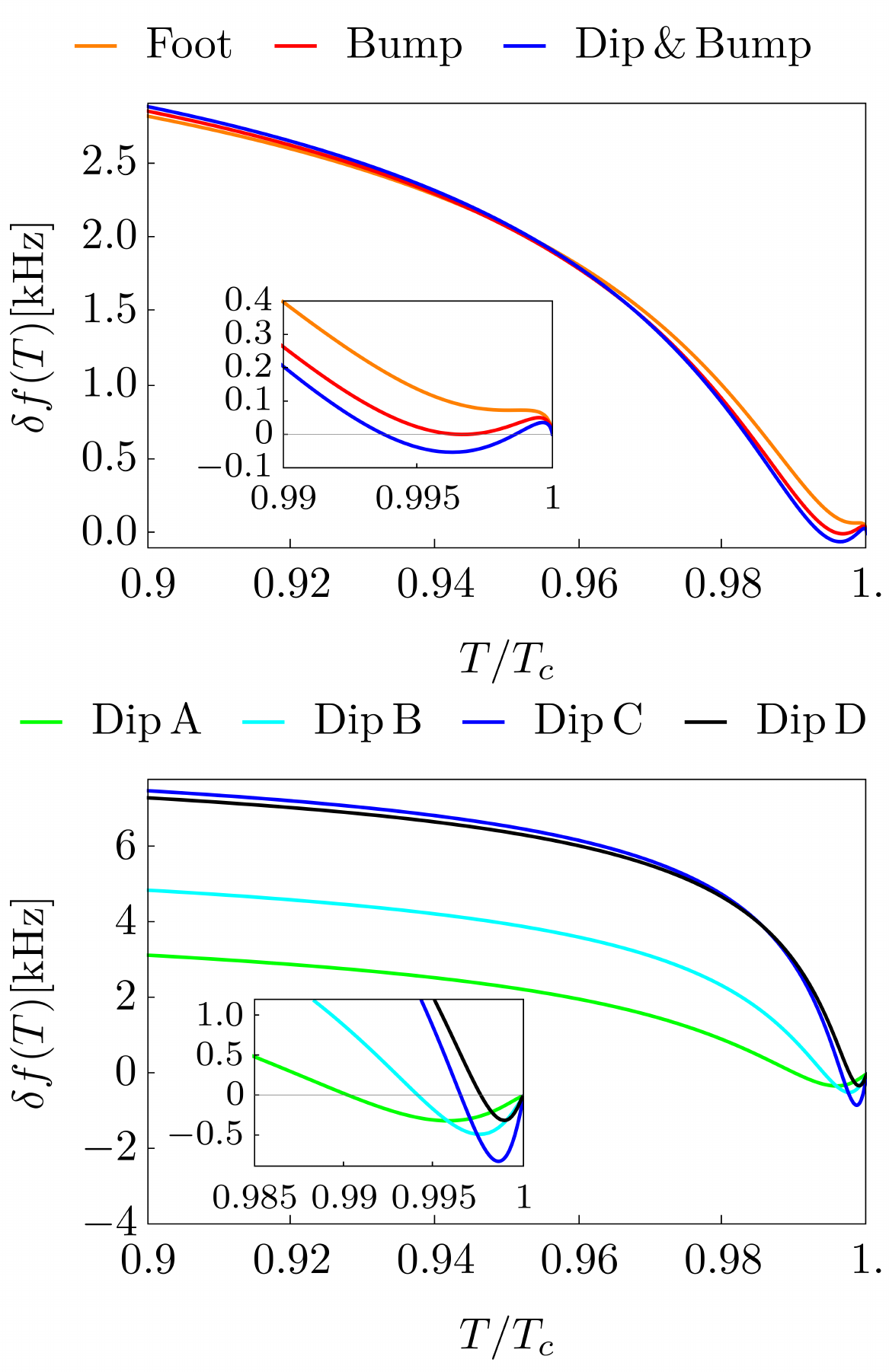}
    \caption{Behavior of $\delta f(T)$ in the vicinity of $T_c$, considering different values of $\Gamma$ and $\Gamma_s$ scattering constants.}
    \label{fig:FreqShift_Regimes}
\end{figure}

In this section, we present and discuss the results of the numerical calculations for the frequency shift based on the DS theory. We combine Eq.~\eqref{eq:ResFreqShift} with Eq.~\eqref{eq:Sigma_Num}, assuming local limit behavior in superconducting and in normal state, according to the discussion from Sec.~\ref{sec:Temp_Range}. The most interesting message can be conveyed by explaining Fig.~\ref{fig:FreqShift_Regimes}  together with Tab.~\ref{tab:Regimes} containing relevant parameters used in the numerics.

\begin{center}
\begin{table}[h!]
\begin{tabular}{ |c|c|c|c|c|c|c| }
 \hline
 & $\Gamma[\omega]$ & $\Gamma_s[\omega]$ & Regime & $\ell[\SI{}{nm}]$ & $X_n[\SI{}{m\Omega}]$ & $\tilde{f}[\SI{}{kHz}]$\\
 \hline
 Foot & $0.33$ & $26$ & $\Gamma \lessapprox \omega \ll \Gamma_s \ll \Delta_0$ & $697$ & $2.1$ & $4.9$\\
 \hline
 Bump & $1.05$ & $26$ & $\Gamma \approx \omega \ll \Gamma_s \ll \Delta_0$ & $679$ & $2.1$ & $5$\\
 \hline
 D\&B & $1.5$ & $26$ & $\omega \lessapprox \Gamma \ll \Gamma_s \ll \Delta_0$ & $668$ & $2.1$ & $5.1$\\
 \hline
 Dip A & $5$ & $26$ & $\omega \lesssim \Gamma < \Gamma_s \ll \Delta_0$ & $592$ & $2.2$ & $5.4$\\
 \hline
 Dip B & $5$ & $50$ & $\omega \lesssim \Gamma \ll \Gamma_s \lesssim \Delta_0$ & $334$ & $3$ & $7.1$\\
 \hline
  Dip C & $5$ & $100$ & $\omega \lesssim \Gamma \ll \Gamma_s \lesssim \Delta_0$ & $175$ & $4.1$ & $9.9$\\
 \hline
 Dip D & $0.33$ & $100$ & $\Gamma \lessapprox \omega \ll \Gamma_s \lesssim \Delta_0$ & $183$ & $4$ & $9.7$\\
 \hline
\end{tabular}
\caption{Individual regimes together with its considered scattering constants $\Gamma$ and $\Gamma_s$ in units of the angular resonant frequency $\hbar\omega$ (for convenience $\hbar=1$, and $f=\SI{1.3}{GHz}$). For each regime, we also add the parameter inequalities as well as calculated values of mean free path $\ell$, normal state reactance $X_n$, and the scale $\tilde{f}$.}
\label{tab:Regimes}
\end{table}
\end{center}

The SRF Nb-motivated values $\hbar\omega=\SI{5.4}{\mu eV}$ and $\Delta_0~=~\SI{2}{meV}\approx370\hbar\omega$ correspond to the assumed values of the resonant frequency and superconducting gap. As can be seen, we choose the externally fixed value of $\hbar\omega$ as the convenient energy reference value. The mean free path values are calculated as $\ell=\hbar v_F/(2\Gamma_n)$. Values of $X_n$ are calculated from Eq.~\eqref{eq:Rn_or_X_n}, since $\omega\tau \ll 1$ for all of the considered cases.

In the following, we discuss our findings concerning the experimentally seen regimes in Ref.~\cite{Bafia21}
and calculations incorporating anisotropy of the superconducting gap and inhomogeneous disorder in the screening region of SRF cavities, presented in Ref.~\cite{Ueki22}.

{\it i)} In our perspective, the individual regimes displayed in Fig.~\ref{fig:FreqShift_Regimes} correspond to different combinations of considered values of the pair-breaking and pair-conserving scattering rates within the DS homogenous theory approach.

{\it ii)} We compare our results (focusing mainly on the insets in Fig.~\ref{fig:FreqShift_Regimes}) with the observed features reported in Fig.~1 of Ref.~\cite{Bafia21} (or Fig.~6.1 of Ref.~\cite{Bafia20}). 
Let us look at the first four regimes from Tab.~\ref{tab:Regimes} with $\ell~\lesssim~\SI{700}{nm}$, reminding us the cleaner of the two cases discussed in Sec.~\ref{sec:Temp_Range}.
Our results are in the correct ranges of temperatures~$T\in(0.99,1)\,T_c$ and resonant frequency shift differences $\delta f\in(-0.3, 0.5)\,\SI{}{kHz}$ scale, assuming $\tilde{f}~\approx~\SI{5}{kHz}$, coming from $f~=~\SI{1.3}{GHz}$, $G~=~\SI{270}{\Omega}$, and $X_n~\approx~\SI{2}{m\Omega}$. 
\begin{figure}[h!]
    \includegraphics[width = 0.45\textwidth]{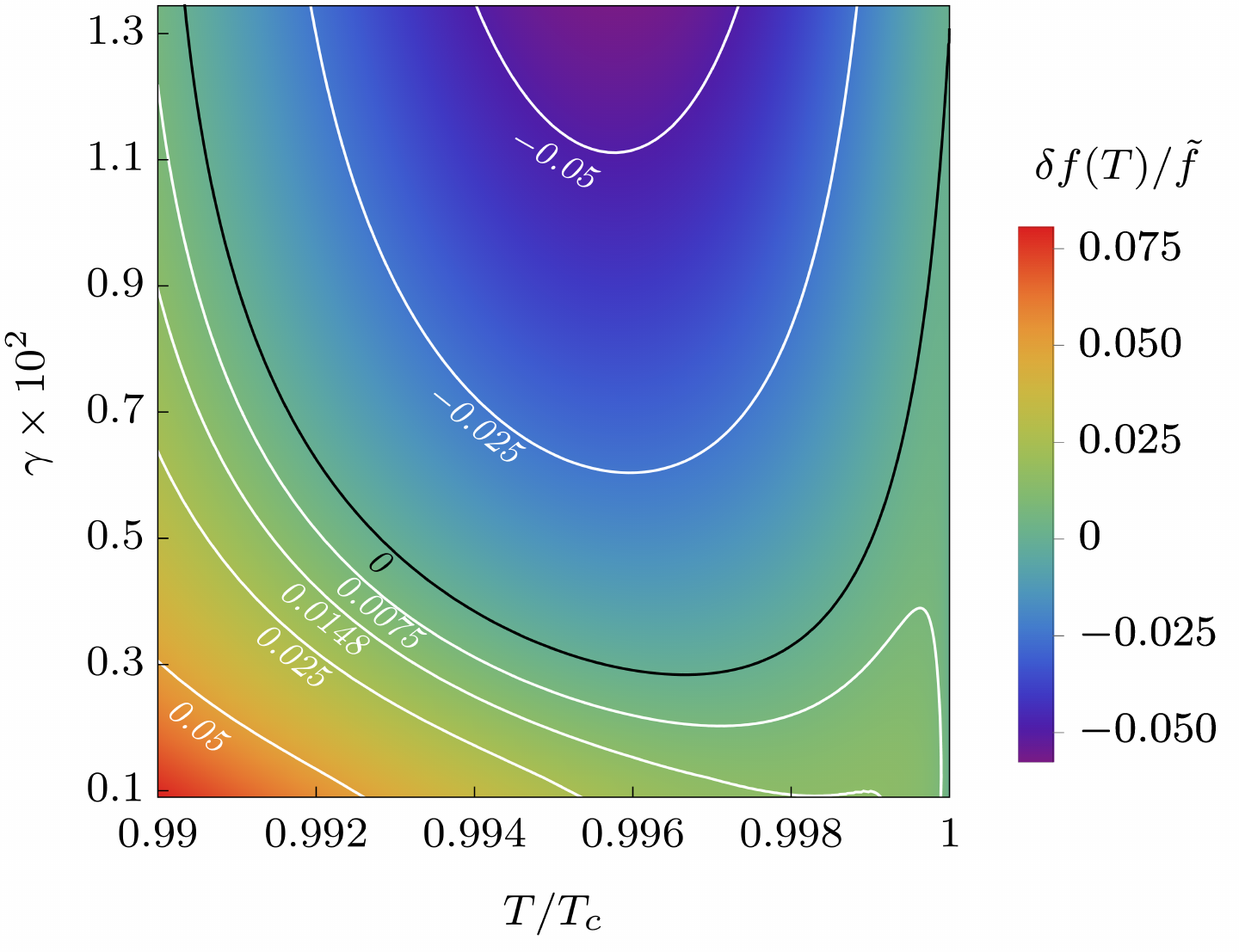}
    \caption{
    Density and contour plot of $\delta f(T)/\tilde{f}$ considering regime of large $\ell$ from Tab.~\ref{tab:Regimes}. We assume
    $\gamma_s = 26 \hbar \omega/\Delta_0 \approx 0.07$, where $\Delta_0=\SI{2}{meV}$, and $\gamma\in (0.33, 5)\times\hbar\omega/\Delta_0\approx\langle 0.1,1.3\rangle\times10^{-2}$.
    }
    \label{fig:Contours_Regimes}
\end{figure}
In Fig.~\ref{fig:Contours_Regimes} we depict different regimes from Fig.~\ref{fig:FreqShift_Regimes} and Tab.~\ref{tab:Regimes} in the density plot of $\delta f(T)/\tilde{f}$, plotted as a continuous function of $\gamma$, considering fixed value of $\gamma_s$. The regimes beginning with the {\it Foot} and ending with the pure {\it Dip} signature can be understood from the shape of the individual contours.

{\it iii)} Next, in the comparison with the theory\footnote{To be more specific, we compare the red curve in Fig.~2 of Ref.~\cite{Ueki22}, using their set of relevant parameters $\tau T_{c_0}=3.29$, $T_{c_0}=\SI{9}{K}$, and $v_F=\SI{0.26}{Mm/s}$.} from Ref.~\cite{Ueki22}, we work on the similar scale of $R_n\sim\SI{2}{m\Omega}$ and $\ell\sim\SI{700}{nm}$.
    
{\it iv)} Note also the increasing depth in Dip A, B, and C with the decreasing $\ell$, agreeing with the experimental findings, reported in Fig.~2(b), and Fig.~2(c) of Ref.~\cite{Bafia21} (or Fig.~6.3b, Tab.~6.2, and/or Fig.~6.7a in Ref.~\cite{Bafia20}). Our DS theory-based analysis explains this trend by the increasing values of the scattering rates. They decrease\footnote{Mainly, the increasing value of $\Gamma_s$ dominates.} $\ell$ and increase the dip depth in $\delta f(T)$.
    
{\it v)} Continuing the idea from the previous point, we explain the decrease of $T_c$ with decreasing $\ell$ (again in agreement with the experimental findings in Fig.~2(c) of Ref.~\cite{Bafia21} or Fig.~6.6 of Ref.~\cite{Bafia20}). We naturally expect that with increasing, usually dominant $\Gamma_s$ (suppressing~$\ell$), the pair-breaking $\Gamma$ increases too. Consequently, increasing $\Gamma$ suppresses $T_c$ \cite{Herman16}.

{\it vi)} For completeness, we also show the regime of the Dip D in Fig.~\ref{fig:FreqShift_Regimes}. As can be seen in Tab.~\ref{tab:Regimes}, this regime represents an unnatural combination of the relatively small $\Gamma$ and yet relatively large $\Gamma_s$, which may be experimentally difficult to achieve.

{\it vii)} From Sec.~\ref{sec:Temp_Range} we see that adding more disorder (reducing $\ell$) ensures local response to the electromagnetic field on both sides of the superconductor-normal metal transition. From Tab.~\ref{tab:Regimes} we see that decreasing $\ell$ results in the {\it Dip} signature of the $\delta f(T)$. Therefore, we expect this regime to often appear in the dirty samples. Someone could ask whether there is an even simpler effective description of this disorder-driven phenomenon. In Appendix~\ref{app:Dip_2FM} we show that even when we omit the difference between pair-conserving and pair-breaking disorder, we can still describe the basic characteristics of the {\it Dip} signature within a very simple two-fluid model approach. The provided analysis is suitable, for example, for explanation purposes or scattering time order of magnitude estimates without the need for more sophisticated tools. It also reveals the role of the ratio of the superconducting and normal state electromagnetic field penetration depths.

{\it viii)} As can be understood from the experimental results reported in Fig.~2(b) of Ref.~\cite{Bafia21} (or Ref.~\cite{Bafia20}, Fig.~6.3b), the {\it standard} regime appears in the clean samples with relatively large $\ell$ (with relatively large uncertainty) compared with other samples. Our estimates suggest that such a regime might already combine local superconducting and strongly anomalous response close to $T_c$, according to the analysis in Sec.~\ref{sec:Temp_Range}. Therefore, we do not address this regime in our currently presented calculations.

\subsection{Analytical considerations}\label{Sec:DipAnalysis}

One of the advantages of our approach is the ability to calculate the behavior in the vicinity of $T_c$ also analytically. In the following paragraphs and Appendix:~\ref{Appendix:ideal dirty} we do so, considering several regimes, interesting also from the experimental point of view. Note that in the following analysis, we omitted the strong coupling corrections for simplicity. These corrections may (as may be seen later in the text) slightly shift our results and may be needed for direct comparison with the experiment focused on niobium-based materials \cite{Herman21}.

{\it (Moderately clean regime)} In the following, we focus on behavior of $\delta f(T\rightarrow T_c)$ in the experimentally very relevant regime of the considered parameters $\hbar\omega~<~\Gamma~\ll~\Gamma_s~\lesssim ~\Delta_0$. The optical conductivity can be in this regime linearized in the arbitrarily small parameter $\Theta \equiv 1 - T/T_c$ as \cite{Herman21}
\begin{equation}\label{eq:sigma_mod_clean_ideal}
    \sigma_s(T)/\sigma_0 = 1 + \big(f_1(\Gamma, \Gamma_s) + i f_2(\Gamma, \Gamma_s, \omega)\big)\Theta.
\end{equation}
Considering the most interesting, ideal scenario ($\Gamma~\ll~T_c~\sim~\Delta_{00}$)  \cite{Herman21}
\begin{equation}\label{eq:f_mod_clean_ideal}
    f_1 = \frac{\gamma_{\sigma}}{4}\frac{\Gamma_s}{\Gamma_n}\frac{\Delta_{00}}{\Gamma},\quad f_2 = \frac{2\Gamma_n}{\omega}\frac{n_s(0)}{n}A = \left(\frac{\delta}{\lambda_{L}(0)}\right)^2\frac{A}{2},
\end{equation}
where we defined the constant $\gamma_{\sigma}=4\gamma_e\pi^2/(7\zeta(3))\approx 8.356$. The number $\gamma_e = e^{E}\approx1.781$, since $E \approx 0.577$ is the Euler-Mascheroni constant.

The auxiliary factor $f_1$ is straightforward; meanwhile, $f_2$ deserves further commentary. We utilize superfluid fraction $n_s(0)/n$ at $T=\SI{0}{K}$ from Appendix~\ref{Appendix:App_sc_frac}, given by Eq.~\eqref{eq:dens_frac_1} (Eq.~\eqref{eq:dens_frac} in general) found in Ref.~\cite{Herman17b, Herman21}. Next, we neglect the weak dependence of $A\approx 2.2$ on the disorder in a very broad range of considered scattering rates \cite{Herman21}. The resulting $\delta f(T)$ close to $T_c$ from Eq.~\eqref{eq:ResFreqShift} acquires elegant form
\begin{equation}\label{Eq:Slope}
\delta f(T)/\tilde{f} = \left(f_1 - f_2\right)\Theta/2,
\end{equation}
immediately suggesting the possibility of different signs for the slope defined as $\delta f'(\Theta)\equiv \partial \delta f(\Theta)/\partial \Theta$ at $\Theta = 0$. Note that once we ignore the weakly disorder-dependent numeric prefactors and assume $\Gamma_n\approx\Gamma_s$, the difference of $f_1-f_2$ can be understood as a comparison of the $1/$relative pair breaking (with respect to the gap) and the second power of the ratio of skin-effect and zero-temperature superconductor penetration depths.

\begin{figure}[h!]
    \includegraphics[width = 0.43\textwidth]{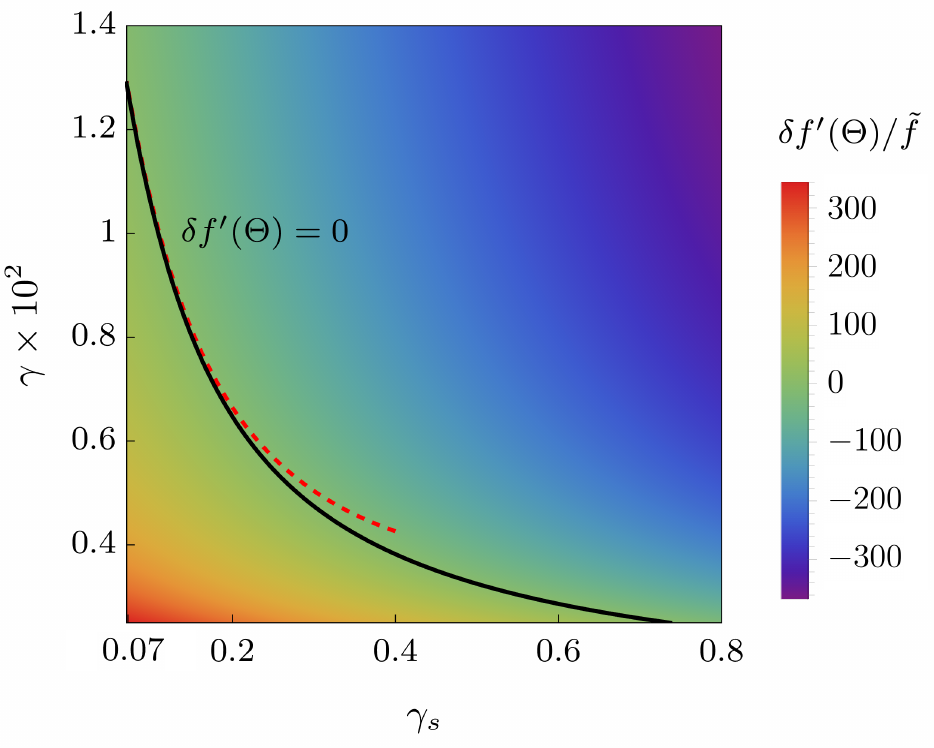}
    \caption{Sign change of the slope $\delta f'(\Theta)/\tilde{f}$ from Eq.~\eqref{Eq:Slope} in the relevant range of considered scattering rates $\gamma$ and $\gamma_s$. Angular frequency corresponds to $\omega=2\pi\times\SI{1.3}{GHz}$.}
    \label{Fig:Slope_mod_clean}
\end{figure}

In Fig.~\ref{Fig:Slope_mod_clean} we depict the sign change in the most interesting parameter range. The red dashed contour corresponds to the linearized form of Eq.~\eqref{eq:dens_frac_1} $n_s(0)/n\approx 1-\gamma-\gamma_s\pi/4$ in scattering rates used in $f_2$. The lower boundary of $\Gamma=\hbar\omega$ is defined by the assumptions of the moderately clean regime \cite{Herman21}. The lower boundary of $\Gamma_s$ is defined by the highest considered $\ell$ from Tab.~\ref{tab:Regimes}.

To sum up this part, the most notable fact is that the simple mean-field theory of Dynes superconductors, assuming homogeneous pair-breaking and pair-conserving disorder, explains both kinds of slope signs in a moderately clean regime. Figure~\ref{Fig:Slope_mod_clean} also places the positive slope in $\Theta$ (negative in $T$) to the {\it clean corner} of the moderately clean regime, suggesting relatively high $\ell$. This result is in (at least) qualitative agreement with the trends seen in the recent experimental findings \cite{Bafia21}. 

\begin{figure}[h!]
    \includegraphics[width = 0.435\textwidth]{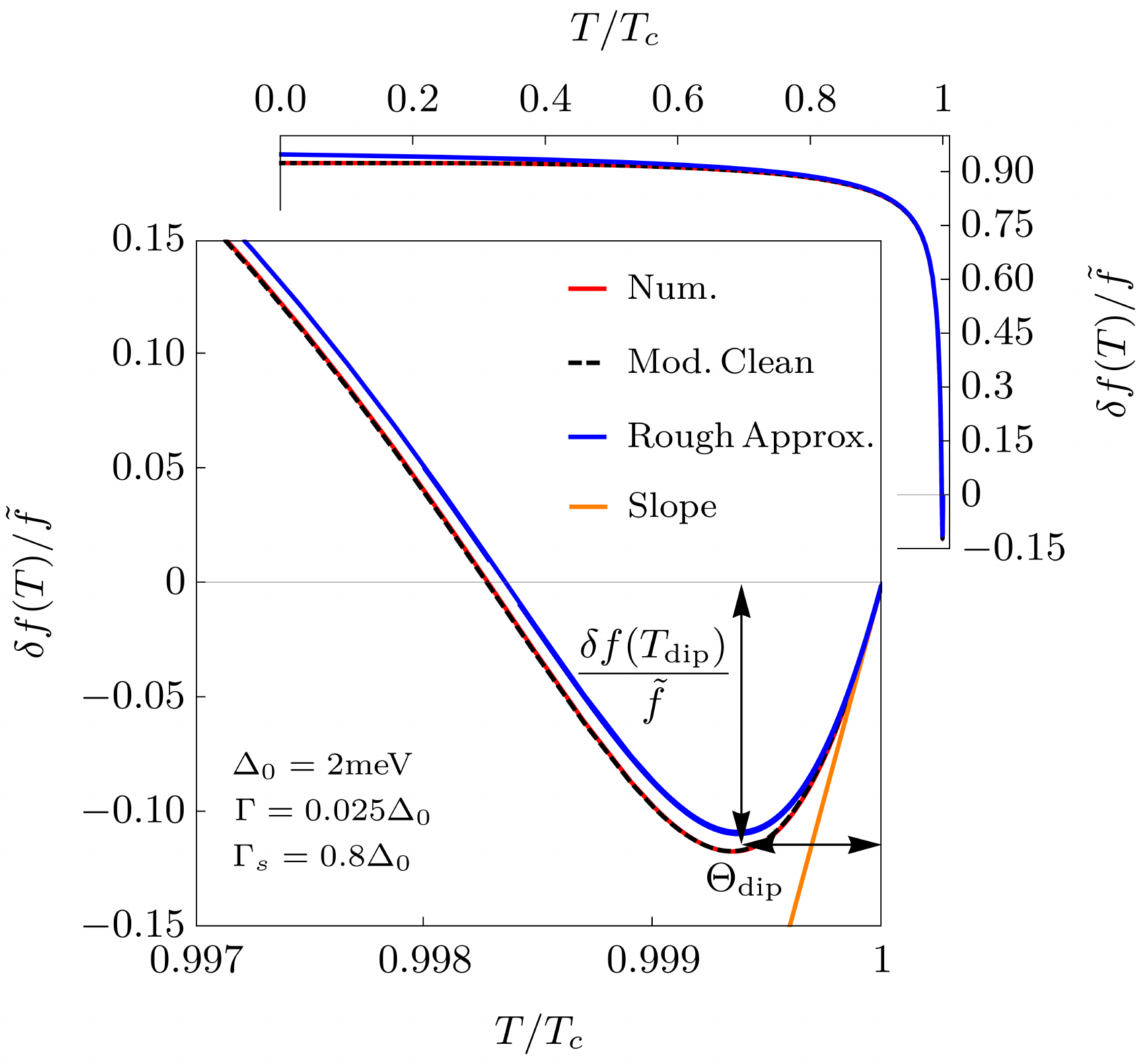}
    \caption{Dip in the moderately clean regime of DS model. Used parameters result in $\omega\tau\approx 2\times10^{-3}$ and $f_1/f_2\approx0.1$.}
    \label{fig:ModCleanIdealAnalytics}
\end{figure}

Focusing, however, in the other direction, pointing to the {\it dirty corner} region of Fig.~\ref{Fig:Slope_mod_clean}, with the clear dip signature, or $\delta f'(\Theta)<0$, it is easy to note that the ratio (using factors from Eq.~\ref{eq:f_mod_clean_ideal} and assuming $\Delta_0=\SI{2}{meV}$ and $f=\SI{1.3}{GHz}$)
\begin{equation*}
    \frac{f_1}{f_2}=\frac{\gamma_{\sigma}}{4A}\frac{\Gamma_s}{\Gamma_n}\frac{\Delta_{00}}{\Gamma}\frac{n \omega\tau}{n_s(0)}=\frac{\gamma_{\sigma}}{2A}\frac{\Gamma_s}{\Gamma_n}\frac{\Delta_{00}}{\Gamma}\left(\frac{\lambda_{L}(0)}{\delta}\right)^2
\end{equation*}
decreases with increasing scattering rates and $f_1/f_2\lesssim0.1$. In addition, $\omega\tau\ll1$ naturally decreases with increasing scattering. In such a case, a simple and rough approximation can be constructed by exploiting Eq.~\eqref{eq:sigma_mod_clean_ideal} and Eq.~\eqref{eq:f_mod_clean_ideal} within Eq.~\eqref{eq:ResFreqShift}. The result of this approximation is plotted together with the exact numerics in Fig.~\ref{fig:ModCleanIdealAnalytics}.

A very rough estimate of the dip's width can be obtained as a solution of $\delta f (T_w) = 0$ for $T_w \neq T_{c}$, once we neglect $\omega\tau$ compared with $1$ or $\pi/2$ in Eq.~\eqref{eq:ResFreqShift}. The resulting equation reads $\sqrt{2\sigma_0/|\sigma_s(T_w)|}\sin\big(\pi/4+\varphi_s(T_w)/2\big)=1$. Its solution, considering series up to $\Theta^2$, leads to
\begin{align}\label{eq:Theta_w_mod_clean}
\Theta_{w}&\equiv1-\frac{T_{w}}{T_c}=\frac{4(f_1-f_2)}{3(f_1^2-2f_1f_2-f_2^2)},\\
&\approx \frac{4}{3f_2}= \frac{4}{3A}\frac{n \omega\tau}{n_s(0)}=\frac{8}{3A}\left(\frac{\lambda_{L}(0)}{\delta}\right)^2,\nonumber
\end{align}
reveals the role of (small but finite) $\omega\tau$, divided by the superfluid fraction $n_s(0)/n\leq1$, which is suppressed by the disorder. Alternatively, one can interpret $\Theta_w$ throughout the ratio of penetration depths. Note also the linear scaling of the width with the resonant frequency $f$, $\Theta_{w}\propto f$, considering $f_1\ll f_2$. The value of the depth in $\Theta_{\mathrm{dip}}\approx\Theta_{w}/2$ results in 
\begin{align*}
    \frac{\delta f\left(T_\mathrm{dip}\right)}{\tilde{f}}&=\frac{\Theta_{\mathrm{dip}}\delta f'(\Theta)\rvert_{\Theta=0}}{2} =\frac{(f_1-f_2)^2}{6(f_1^2-2f_1f_2-f_2^2)},\\
    &\approx -0.167 + 1.393\frac{\gamma_s}{\gamma_n\gamma}\frac{n}{n_s(0)}\frac{\omega\tau}{A},\\
    &\approx -0.167 + \frac{2.786}{A}\frac{\gamma_s}{\gamma_n\gamma}\left(\frac{\lambda_{L}(0)}{\delta}\right)^2.
\end{align*}
Note that in case $f_1\ll f_2$, our approximation shows a natural scale of the dip depth $\delta f\left(T_\mathrm{dip}\right)\approx -0.17\tilde{f}$.

In Appendix:~\ref{Appendix:ideal dirty} we also present results of the analysis in the ideal dirty limit, in other words, the regime when $\hbar\omega\ll\Gamma\ll\Delta(0) \approx \Delta_{00}\ll\Gamma_s$. It is fair to say that within such a limit, our results require a superconducting material with a highly isotropic superconducting order parameter, since the anisotropy will cause a reduction of $T_c$, moving away from our considered assumptions. However, the found scale of $\delta f\left(T_\mathrm{dip}\right)~\sim-0.17\tilde{f}$ is shown to be present in the numerical solution even for larger $\gamma \sim 0.1$, corresponding to the noticeable $T_c$ reduction in our approach.

{\it (Large pair-breaking)} The last statement is related to the following question. What happens in the regime of strong pair breaking, when $\Gamma\gg T_c$ with arbitrary $\Gamma_s$? Factor $f_1$ changes to~\cite{Herman21, Lebedeva24}
\begin{equation}\label{eq:f1_mod_clean_large_gamma}
    f_1 = -\frac{2\gamma_s + 3\gamma}{\gamma_s + \gamma}\frac{1}{\gamma^2} <0.
\end{equation}
This implies directly $\delta f'(\theta)<0$ in Eq.~\eqref{Eq:Slope} because, generally $f_2>0$. This suggests a {\it Dip} behavior in the considered regime. We may also note that $f_1$ is highly suppressed by $\gamma\gg 1$, leading to $f_1/f_2\rightarrow 0$ leaving $\Theta_w$ in the form of Eq.~\eqref{eq:Theta_w_mod_clean} and the scale of $\delta f\left(T_\mathrm{dip}\right)\approx -0.17\tilde{f}$.

\begin{figure}[h!]
    \includegraphics[width = 0.4\textwidth]{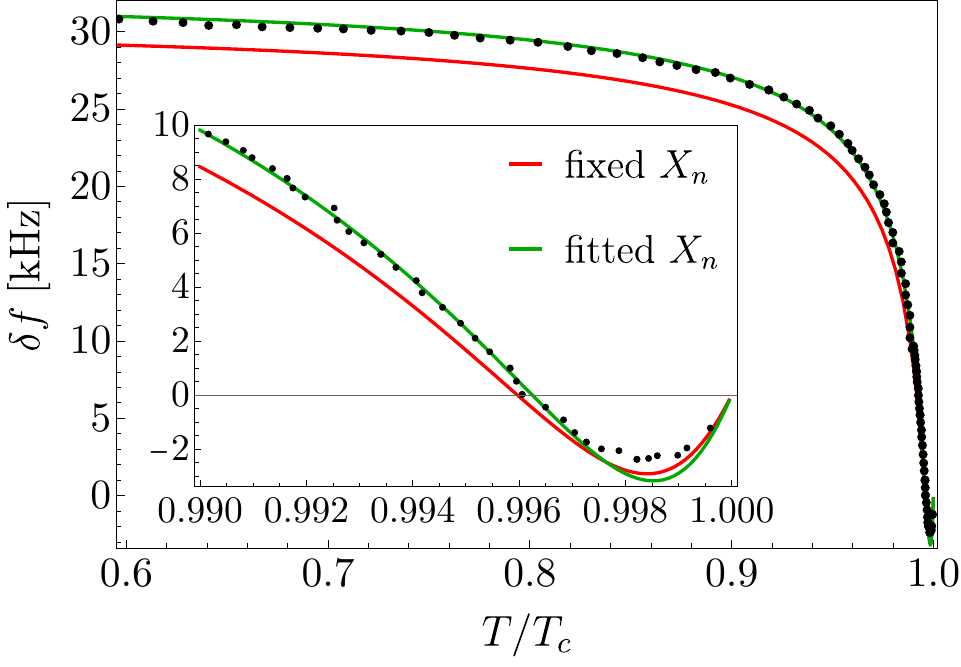}
    \caption{Considered parameters: resonant frequency $f~=~\SI{2.6}{GHz}$, $\Delta_0 = \SI{1.55}{meV}$ \cite{Zarea23a}, pair-conserving scattering rate corresponding to fixed~ $X_n~=~\SI{7.1}{m\Omega}$ (fitted~ $X_n~=~\SI{7.5}{m\Omega}$) is $\gamma_s = 0.52$ ($\gamma_s=0.58$), and resulting $\ell = \SI{121}{nm}$ ($\ell = \SI{108}{nm}$).}
    \label{Fig:comparison_and_fit}
\end{figure}

\subsection{Comparison and fitting of the experimental dip}

In Fig.~\ref{Fig:comparison_and_fit}, we compare and also fit the dip in $\delta f(T)$ on the data elaborated in Ref.~\cite{Zarea23a} (their Fig.~2 and Fig.~3) using the DS theory. For both procedures we first determine the pair-breaking rate $\gamma = 0.026$ from the suppressed $T_c =\SI{9}{K}$ to $T_{c0} =\SI{9.33}{K}$ ratio on the considered N-doped Nb sample \cite{Zarea23a}. Note that in the regime of small (linear) $\gamma\ll1$ we can generally use \cite{Lebedeva24, Tinkham392}
\begin{equation}\label{eq:gamma_lin}
    \gamma = \gamma_e (2/\pi)^2 (1 - T_c/T_{c0}). 
\end{equation}

We also phenomenologically include the influence of strong coupling corrections by rescaling the temperature dependence $\Delta(T)$ by the factor $x$, reflecting the deviation of the $\Delta_{00}/T_{c0}$ ratio from the BCS value in pure Nb. We proceed in a similar way as described in more detail in Ref.~\cite{Herman21}. In our case, factor $x = 1.145$ is determined from the experimentally found ratio $2\Delta_{00}/T_{c0}=4.04$ on pure Nb samples\footnote{Adjusting the strong-coupling correction parameter $x$ to the ratio $\Delta_0/T_c$ for our specific sample does not lead to any substantial changes.} from Ref.~\cite{Groll18}.

Next, in the comparison scenario, we utilize the experimentally determined value of $X_n = R_n = \SI{7.1}{m\Omega}$. This value together with the rest of the considered parameters and Eq.~\eqref{eq:Rn_or_X_n} unambiguously fixes the value of $\gamma_s$. It means that we have no free parameters. The resulting function describing $\delta f(T)$ is plotted by the red curve in Fig.~\ref{Fig:comparison_and_fit}. For control purposes, in a fitting scenario, we consider $\gamma_s$ as a free parameter which we determine by the least squares method considering experimental data and DS theory. The resulting curve is plotted by the green line in Fig.~\ref{Fig:comparison_and_fit}.

The overall result reveals the following message. First, we find the agreement of the experiment and both procedures to be satisfactory, especially if we focus on the width of the dip. Second, the fitted value of $X_n$ deviates on the level of $5\%$ from the experimentally determined value. This means that our fit does not contradict the independent experimental measurement of the normal state surface reactance. One way or another, both procedures point toward the moderately clean regime behavior described within the DS theory.

In Appendix:~\ref{Appendix:Dip_Other_freq} we compare all four cavities with different resonant frequencies from Ref.~\cite{Ueki22}. The analysis points toward very similar values of the gap $\Delta_0$ and the scattering constants $\Gamma$ and $\Gamma_s$, considering all four cases. It even shows the trend of continual proportion between $\Gamma_s$ and $\Gamma$, where always $\Gamma/\Gamma_s\lesssim5\%$. This agrees with the expected behavior of increasing pair-breaking naturally following increasing dominant pair-conserving scattering.

\begin{figure}[h!]
    \includegraphics[width = 0.48\textwidth]{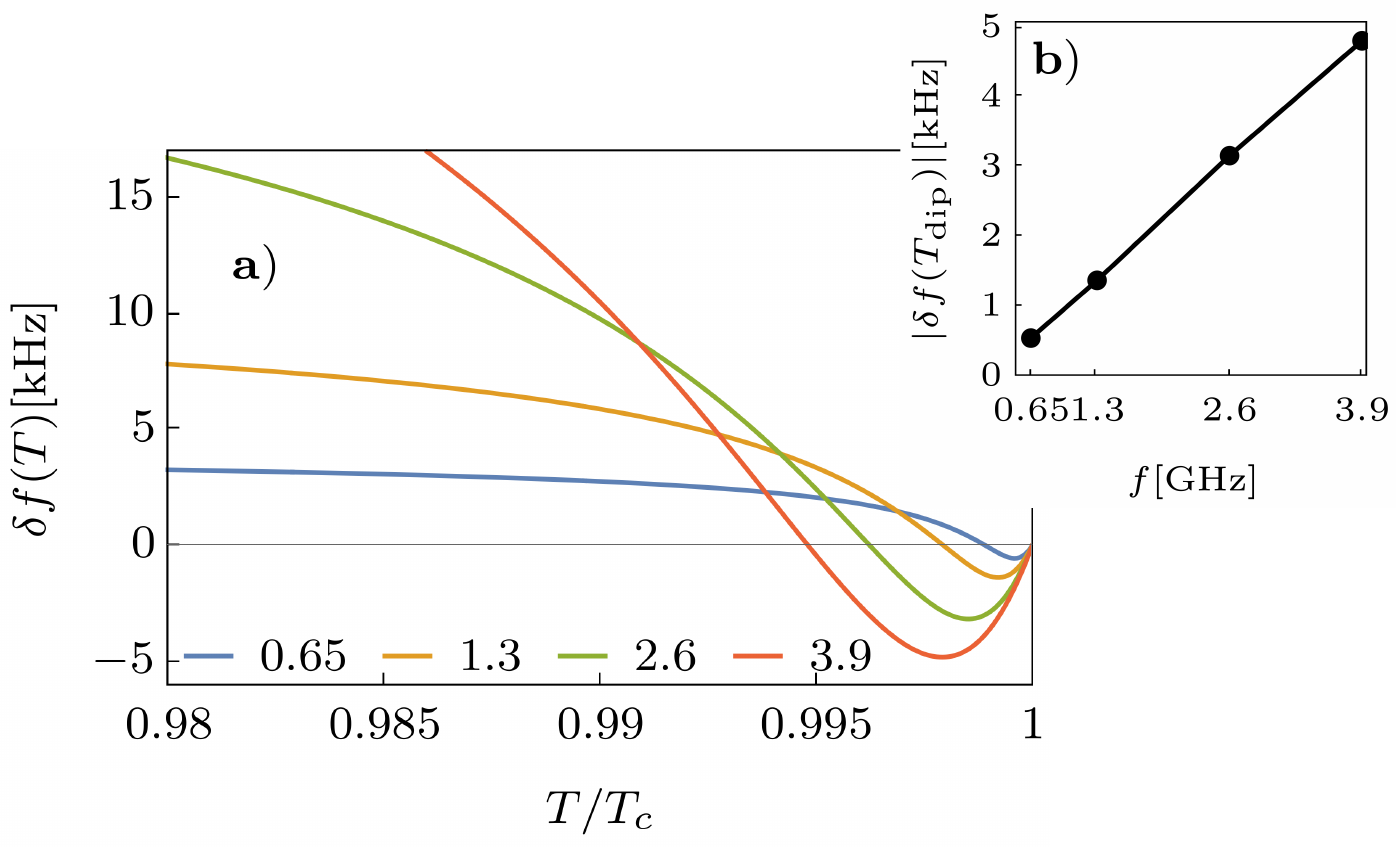}
    \caption{Resonant frequency dependence of $\delta f(T)$ considering values of $f=\{ 0.65, 1.3, 2.6, 3.9\}\SI{}{GHz}$. Considered scattering rates are $\gamma = 0.026$ and $\gamma_s=0.58$.}
    \label{Fig:freq}
\end{figure}

Since the analysis provided in Appendix:~\ref{Appendix:Dip_Other_freq} indicates very similar properties of the utilized superconducting niobium, for completeness, we also show the resonant frequency dependence of the fitted dip from Fig.~\ref{Fig:comparison_and_fit} in Fig.~\ref{Fig:freq}. The considered $\ell\sim \SI{100}{nm}$ is similar to measurements shown in Fig.~3(b) of Ref.~\cite{Bafia21} (or Fig.~6.8 in Ref.~\cite{Bafia20}). Note the overall trend shown in Fig.~\ref{Fig:freq} {\bf a)} as well as the almost linear scaling of the depth of the dip with the resonant frequency shown in Fig.~\ref{Fig:freq} {\bf b)}, being in agreement with the experimental observations. 

In principle, considering the lowest resonant frequency of $f=\SI{0.65}{GHz}$, the dip can be described within the analytical approach introduced in the section focusing on a moderately clean regime, since the required inequalities are fulfilled. For curiosity, the resulting ratio $f_1/f_2\approx0.07$, $\Theta_w\approx10^{-3}$ and $\delta f(T_{\mathrm{dip}})\approx\SI{-0.56}{kHz}$, is in good agreement with the numerics shown in Fig.~\ref{Fig:freq} {\bf a}) and {\bf b}). However, as the increasing resonant frequency gets closer to the considered scale of $\Gamma$, it violates the assumptions under which Eq.~\eqref{eq:sigma_mod_clean_ideal} and Eq.~\eqref{eq:f_mod_clean_ideal} were derived, the analytical approach cannot be used, and we are left with numerics.

\section{Quality}\label{sec:Q}

In a recently published study, the authors analyze the quality factor using the solution of the nonequilibrium Eilenberger equations \cite{Kubo22}, assuming the presence of the pair-breaking and pair-conserving disorder, similar to our considerations. We aim to extend their analysis. Alternatively, the authors of Refs.~\cite{Ueki22, Zarea23b} include the effect of the superconducting gap anisotropy, disorder inhomogeneity, and strong electron-phonon coupling within the Keldysh formalism \cite{Rainer95}. Both approaches use quasiclassical Green's functions, where the one-particle Green's functions are integrated over the band structure energies. Within our straightforward and mostly analytical model describing the DS, we consider homogeneous pairing and disorder scattering interactions. We also use the integrated Green function; however, its form is motivated by the local solution of the self-consistent CPA equations. In addition, we study only the surface resistance part of the quality factor $Q_s$, neglecting other effects coming from two-level systems \cite{Muller19}, nonequilibrium quasiparticles \cite{Visser14}, and trapped vortices \cite{Gurevich17} contributing to the overall internal quality factor.

Realizing $Q_s = G/R_s$, exploiting Eq.~\eqref{eq:Rs} and noting that in anomalous and also local limit $k<0$ leads to
\begin{equation}
    \frac{Q_s}{Q_n} = \left(\frac{|\sigma_s|}{|\sigma_n|}\right)^{|k|}\frac{\cos(\alpha_n)}{\cos(\alpha_n + |k|\delta\varphi)}\label{eq:Qs},
\end{equation}
where $Q_n = G/R_n$. To get a better notion about the relevant numbers in the normal state, we can use values for Dip B in Tab.~\ref{tab:Regimes} and (locally absolutely correct) assumption $X_n = R_n$, resulting in $Q_n\approx 9.1\times 10^4$. Next, focusing on the superconducting state, Eq.~\eqref{eq:Qs} reveals the rapid increase of $Q_s$ once $T$ is decreasing under $T_c$, regardless of the specific model for conductivity, because of the following. i) The ratio of amplitudes will increase with lowering the temperature under the $T_c$ due to the dominating imaginary part $\sigma''_s$ in the superconducting state. ii) For the same reasons (however with the greater effect), the ratio of cosines is much larger than one, since $\delta\varphi \xrightarrow{\sim} \pi/2$ as we go deeply in superconducting region and $(\alpha_n + |k|\delta\varphi) \xrightarrow{\sim} \pi/2$ (assuming $\omega\tau\rightarrow 0$) for both, local and anomalous limit. 

\begin{figure}[h!]
    \includegraphics[width = 0.46\textwidth]{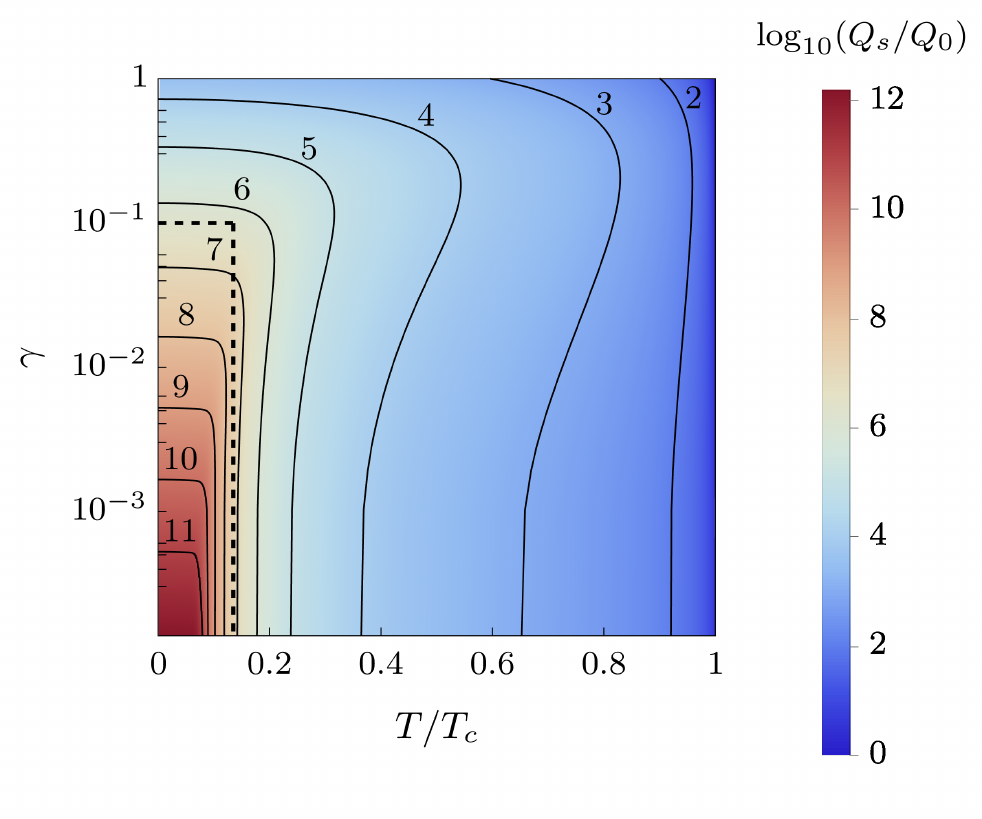}
    \caption{DS quality factor dependence on pair-breaking scattering and temperature for $\gamma_s = 50\hbar\omega/\Delta_{00}$, where $f=\SI{1.3}{GHz}$ and $\Delta_{00}=\SI{2}{meV}$.}
    \label{fig:Qs_color_map}
\end{figure}

In the following, we study the temperature evolution of the superconducting quantities in the local state $(k=-1/2)$. Normalizing them on their normal state values is natural, for example, $R_s/R_n$, $X_s/X_n$, and $Q_s/Q_n$. However, these ratios can require an additional explanation if we study their continuous evolution with the increasing or decreasing scattering rates as the normal state may explicitly depend on their values as well. In such a case, it is better to normalize the superconducting values on units explicitly independent of disorder effects related to $\Gamma_n$ explicitly present in $Q_n$. Therefore, in Fig.~\ref{fig:Qs_color_map} we plot the superconducting quality $Q_s$ according to the DS model by evaluating Eq.~\eqref{eq:Qs} in the experimentally convenient units of $Q_0=\sqrt{\gamma_n}Q_n$.

\begin{figure}[h!]
    \includegraphics[width = 0.4\textwidth]{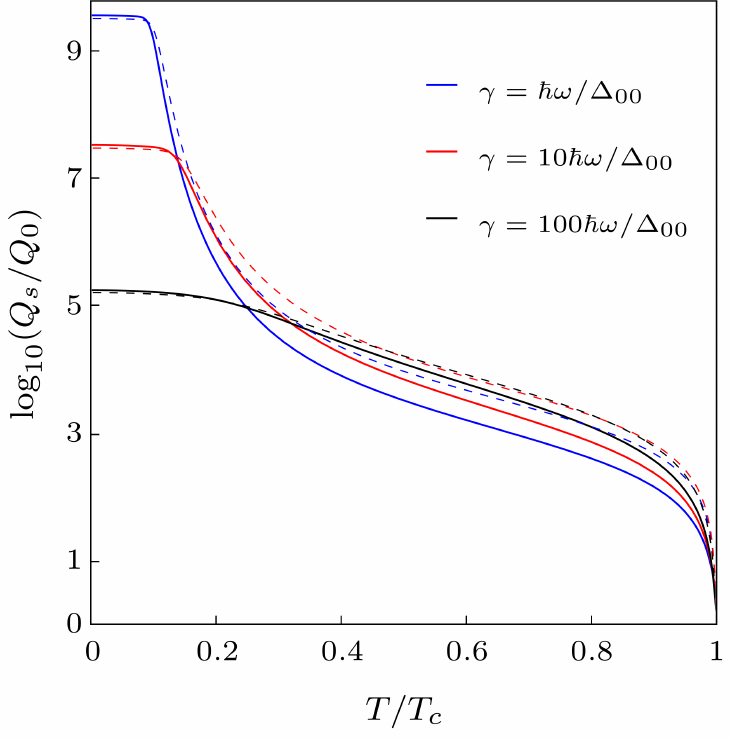}
    \caption{Plot of $Q_s(T)$ for several values of pair-breaking scattering. 
    Full (dashed) lines correspond to $\gamma_s = 50 \hbar\omega/\Delta_{00}\approx 0.13$ ($\gamma_s~\approx~1$).
    }
    \label{fig:Qs_cuts_gamma}
\end{figure}

The $Q_s$ is naturally maximized for the absenting pair-breaking scattering and $T\rightarrow\SI{0}{K}$. Focusing on the lower left corner, surrounding this point\footnote{Highlighted by the black dashed line.} in Fig.~\ref{fig:Qs_color_map}, we can note only weakly decreasing $Q_s$ with $T$ at the low-temperature region (plateaus), assuming the fixed value of $\gamma\lesssim 0.1$. These plateaus, (comparable to regions of almost constant\footnote{Taking into account the logarithmic scale.} surface resistance in Ref.~\cite{Kubo22}) are more clear in Fig.~\ref{fig:Qs_cuts_gamma}, where we plot the cuts considering several chosen values of $\gamma$.

To understand this plateaus, we elaborate on the first interesting temperature scale ($T_p$) and the range of small $\Gamma \ll \Delta_0\approx\Delta_{00}$. This regime includes the experimentally highly relevant moderately clean regime \cite{Herman21}, where it shows to be only weakly dependent on the value of $\gamma_s\lesssim1$, as shown in Fig.~\ref{fig:Qs_cuts_gamma}. The presence of such slowly varying regions of $Q_s$ (mainly at low temperatures) may also be interesting in the context of already introduced frameworks in Refs.~\cite{Zarea23a, Ueki22}.

In our case, $T_p$ corresponds to the rapid decrease of $Q_s(T)$ {\it knee feature} in Fig.~\ref{fig:Qs_cuts_gamma}. To understand this scale, we can analyze Eq.~\eqref{eq:Qs}. Assuming the natural property of the superconducting state at very low temperatures $\sigma_s'(T)/\sigma_s''(T)~\ll~1$ and $\omega\tau\rightarrow 0$ in the normal state, we are left with
\begin{equation}\label{eq:Qs1}
    \left.\frac{Q_s(T)}{Q_n}\right\lvert_{T\ll T_c} = \sqrt{\frac{2}{\sigma_0}}\frac{\sigma_s''(T)^{3/2}}{\sigma_s'(T)},
\end{equation}
immediately showing the role of the real and imaginary parts of the optical conductivity. Next, we exploit that $\sigma''(T)\sim\sigma''(0)$ is still much less temperature dependent than $\sigma'(T)$ at the relevant range of low temperatures. Thus, to proceed further, we express $\sigma'(T)$ straightforwardly in the considered microwave region from Eq.~\eqref{eq:Sigma_Num} as \cite{Herman21}
\begin{equation}\label{eq:sigma_real_T}
    \frac{\sigma_s'(T)}{\sigma_0} = \int_0^{\infty}d\nu\left(-\frac{df(\nu)}{d\nu}\right)\frac{2\Gamma_n}{\varepsilon_2}\left[n_1^2+\frac{\left(p_1\varepsilon_2-p_2\varepsilon_1\right)^2}{|\varepsilon|^2}\right].
\end{equation}
Next, as we are considering low temperatures $T\approx 0.1 T_c$ we may assume $\Delta(T)\approx\Delta_0$, meaning the temperature-relevant effect is coming mainly from the Fermi-Dirac distribution $f(\nu)$. Therefore, assuming $\Gamma\ll\Delta_0\approx \Delta_{00}$, it is enough to investigate the role of Fermi-Dirac distribution at the scale of $\nu\approx\Delta_{00}$. This scale corresponds to the first significant contribution with increasing temperature, since the term in the square brackets in Eq.~\eqref{eq:sigma_real_T} peaks (discussed in the following). Thus, to clarify $T_p$, we analyze 
\begin{align*}
-\left.\frac{df(\nu)}{d\nu}\right\vert_{\nu = \Delta_{00}} &=\frac{1}{4\Delta_{00}x \cosh^2(1/2x)}\approx \frac{e^{-1/x}}{\Delta_{00}x},
\end{align*}
assuming $x=T/\Delta_{00}\ll1$. The exponential starts to cause a steep increase of the considered function (best seen in the log-log scale) around the value $x_p\approx 0.05$, resulting in $T_p/T_{c,0}\approx 0.1$. 

Coming back to the rest of the terms in Eq.~\eqref{eq:sigma_real_T}, we can even note that (for simplicity) in the ideal dirty limit $\Gamma \ll \Delta_{00} \ll \Gamma_s$, the Fermi-Dirac distribution unrelated terms result in
\begin{equation*}
\frac{2\Gamma_n}{\Delta_{00}\varepsilon_2}\left[n_1^2+\frac{(p_1\varepsilon_2-p_2\varepsilon_1)^2}{|\varepsilon|^2}\right]_{\nu=\Delta_{00}}\approx \frac{1}{\Gamma},
\end{equation*}
being independent of $\Gamma_s$ and at the same time highlighting the role of $\Gamma$ in this manner. 

To sum up, $T_p$ is related to the Fermi-Dirac distribution reaching the scale of the underlying superconducting gap, smeared by the pair-breaking scattering. As examined in the dirty limit, it shows to be independent of the pair-conserving scattering. As shown in Fig.~\ref{fig:Qs_cuts_gamma}, with increasing $\gamma$, the temperature scale $T_p$ is first slightly shifted and at the end completely smeared. This is caused by the subgap states and peak broadening (noticeable already only in $n_1^2(\nu)$) for the Fermi-Dirac nonrelated part of Eq.~\eqref{eq:sigma_real_T}. We can also note in Fig.~\ref{fig:Qs_cuts_gamma}, that a higher value of $\gamma_s$ (dashed lines) slightly reduces the $Q_s(T)/Q_0$ in comparison with lower $\gamma_s$ (full lines) considering $T\lesssim T_p$. On the other hand, considering $T\gtrsim T_p$, the effect of larger $\gamma_s$ increases $Q_s(T)/Q_0$ compared with its lower $\gamma_s$ counterpart.

\begin{figure}[h!]
    \includegraphics[width = 0.43\textwidth]{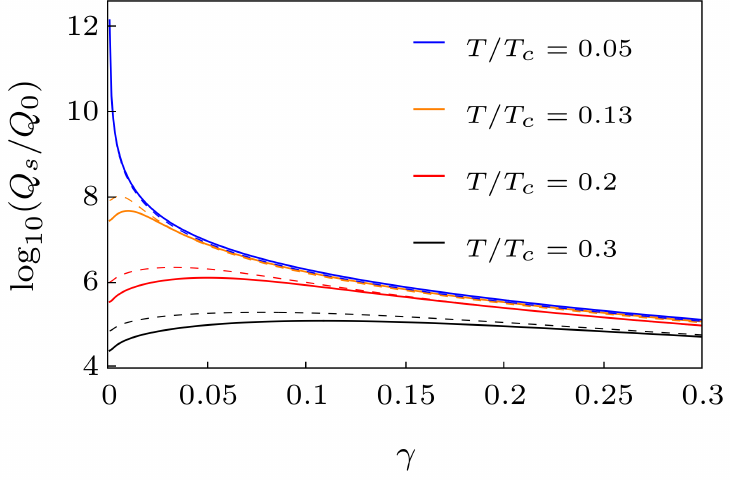}
    \caption{Quality factor as function of $\gamma$ for several values of temperature. 
    Full (dashed) lines correspond to $\gamma_s~=~50 \hbar\omega/\Delta_{00}\approx 0.13$ ($\gamma_s~\approx~1$).
    }
    \label{fig:Qs_cuts_Temp}
\end{figure}

Another interesting feature is the maximum of $Q_s(T)$ at finite, moderately low temperatures  $\omega \ll T \ll T_{c0}$ at a finite value of pair-breaking $\gamma$, plotted in Fig.~\ref{fig:Qs_cuts_Temp}. Such behavior is also present in other approaches, including disorder-related effects \cite{Kubo22, Ueki22, Zarea23a, Gurevich23}. Note that the presence and dependence of this maximum on temperature can also be deduced from the position of the tangent points of vertical tangents (considering fixed $T$) for the individual contours in Fig.~\ref{fig:Qs_color_map}.

\subsection{Quality at zero temperature}

Noting the interesting properties of the $Q_s(T)$ considering range $T\lesssim 0.1 T_c$ in the previous section, let us analytically address the $T \rightarrow \SI{0}{K}$ limit in more detail. In the following, we also assume microwave frequency range $\omega\rightarrow 0$, where \cite{Herman17b, Herman21}
\begin{equation*}
    \frac{\sigma_s''(T)}{\sigma_0} = \frac{2(\Gamma + \Gamma_s)}{\omega}\frac{n_s(T)}{n} =\frac{1}{2}\left(\frac{\delta}{\lambda_L(T)}\right)^2.
\end{equation*}
Combining this equation together with Eq.~\eqref{eq:Qs1}, we are left with
\begin{align}\label{eq:QsT0}
    \frac{Q_s(0)}{Q_n} &= \sqrt{\frac{2}{\sigma_0}}\frac{\sigma_s''(0)^{3/2}}{\sigma_s'(0)} = \frac{4\sigma_0}{\sigma_s'(0)}\left(\frac{n_s(0)}{n}\frac{\Gamma+\Gamma_s}{\omega}\right)^{3/2}.
\end{align}
Alternatively, we can express the residual resistance ratio as 
\begin{equation*}
\frac{R_s(0)}{R_n} = \frac{Q_n}{Q_s(0)}=\frac{2\sigma'_{s}(0)}{\sigma_0}\left(\frac{\lambda_{L}(0)}{\delta}\right)^3,
\end{equation*}
immediately suggesting the interpretation of the absorption ratio multiplied by the ratio of effective cube volumes with side lengths of the corresponding penetration depths. The factor of two can be viewed as a contribution to superconducting absorption coming from two electrons in the Cooper pair.

Once we exploit results for a superfluid fraction $n_s(0)/n$ at $T=\SI{0}{K}$ \cite{Herman17b}, (or Eq.~\eqref{eq:dens_frac})
together with the zero-temperature real part $\sigma_s'(0)$ \cite{Herman21}
\begin{equation}\label{eq:Re_sigma_T_0}
    \frac{\sigma_s'(0)}{\sigma_0} = \frac{\gamma^2}{1+\gamma^2}\times\frac{\gamma_s+\gamma}{\gamma_s+\sqrt{1+\gamma^2}},
\end{equation}
we can study $Q_s(0)$ analytically. In addition, from the role of $\gamma$ ($\gamma_s$) in the first (second) ratio in Eq.~\eqref{eq:Re_sigma_T_0} and role of $\sigma_s'(0)$ in Eq.~\eqref{eq:QsT0} it is clear, why the corresponding residual resistance $\propto 1/Q_s(0)$ is referenced as being subgap-state-induced in the current literature \cite{Kubo22}. Next, using Eq.~(\ref{eq:QsT0},~\ref{eq:dens_frac},~\ref{eq:Re_sigma_T_0}) in the leading order of scattering rates 
$\sigma'_s(0)/\sigma_0\approx\gamma^2(\gamma+\gamma_s)$ and $\sigma''_s(0)/\sigma_0\approx 2(\Gamma+\Gamma_s)/\omega$, we are left with
\begin{equation*}
    \frac{Q_s(0)}{Q_n} = 4\left(\frac{\Delta_0}{\omega}\right)^{3/2}\frac{\sqrt{\gamma+\gamma_s}}{\gamma^2},
\end{equation*}
allowing for quick order-of-magnitude estimates. Assuming a very clean system with the resonant frequency of $f=\SI{1.3}{GHz}$ and $\gamma = \gamma_s \approx 2.7\times 10^{-3}$, being on the same energy scale assuming $\Delta_0=\SI{2}{meV}$, we get $Q_s(0)\sim 10^{8}Q_n$. This demonstrates a huge, yet finite, increase in the quality factor of the superconducting material compared to the normal regime. In Appendix~\ref{Appendix:App_Nb3Sn} we provide a ballpark estimate of pair-breaking contribution to $R_s(0)$ for Nb$_3$Sn.

As can be seen, values of $Q_s(0)$ depend on the specific values of scattering rates. Therefore, we use Eq.~(\ref{eq:QsT0},~\ref{eq:dens_frac},~\ref{eq:Re_sigma_T_0}) to simplify the $Q_s(0)$ in few interesting regimes and generalize the results from Ref.~\cite{Kubo22}. We consider various limits of experimentally accessible scattering rates $\gamma$ and $\gamma_s$ and obtain the following

\begin{widetext}
\begin{equation}\label{eq:Qregimes}
    \frac{Q_s(0)}{Q_0}=\sqrt{2}\left(\frac{\Delta_0}{\omega}\right)^\frac{3}{2}\times 
    \begin{cases}
        \frac{\sqrt{8}}{\gamma^2}\Big[1-\frac{3}{2}\gamma-\left(\frac{3\pi}{8}-1\right)\gamma_s\Big] & \text{if}\quad \gamma \ll 1,\gamma_s \ll 1,\\
        \frac{1}{\sqrt{2(1+\gamma_s)}}\left(\frac{2}{\gamma}\right)^2\left[\left(1+\frac{1}{\gamma_s}\right)\left(\frac{\pi}{2}-\frac{\arccos(\gamma_s)}{\sqrt{1-\gamma_s^2}}\right)-\gamma\right]^{3/2} & \text{if}\quad \gamma \ll \gamma_s \lesssim 1,\\
        \frac{1}{\gamma^2}\sqrt{\frac{\pi^3}{\gamma_s}}\Big[1-\frac{3}{\pi}\gamma-\frac{1}{\gamma_s}\left(\frac{3}{\pi}\ln(\gamma_s)-1\right)\Big] & \text{if}\quad \gamma\ll1\ll\gamma_s,\\
        \sqrt{\frac{8}{\gamma_s\gamma^3}}\Big[1+\frac{\gamma}{\gamma_s}\left(1+\frac{3}{2}\ln\left(\frac{\gamma}{\gamma_s}\right)\right)\Big] & \text{if}\quad 1\ll\gamma\ll\gamma_s,\\
        \frac{1}{\sqrt{\gamma}}\left(1+\frac{\gamma_s}{\gamma}\right)\Big[\frac{2}{\gamma_s}\left(1-\frac{\gamma}{\gamma_s}\ln\left(1+\frac{\gamma_s}{\gamma}\right)\right)\Big]^{3/2} & \text{if}\quad 1\ll\gamma, 1 \ll \gamma_s,\\
        \frac{1}{\gamma^2}\left(1+\frac{\gamma_s}{\gamma}\right) & \text{if}\quad 1\ll\gamma_s\ll\gamma,\\
        \frac{1}{\gamma^2} & \text{if}\quad \gamma_s\ll1\ll\gamma,
    \end{cases}
\end{equation}
\end{widetext}
where $Q_0 = eG\sqrt{n/(\omega m\mu_0\Delta_0)}$ is dependent on the scattering rates only throughout utilizing the experimentally measured gap $\Delta_0 = \Delta_{00}\sqrt{1-2\Gamma/\Delta_{00}}$ at $T = \SI{0}{K}$. Note that in the moderately clean regime (second line) of Eq.~\eqref{eq:Qregimes} the $\gamma_s$-term within the square brackets slightly increases from $1$ for $\gamma_s=0$ to $\pi-2\approx 1.142$ for $\gamma_s~=~1$ and the $1/\sqrt{1+\gamma_s}$ term prevails. Therefore, the zero-temperature quality factor $Q_s(0)$ is much more sensitive to pair-breaking, than the pair-conserving disorder. Even when this behavior is most visible in the low disorder, or ideal clean (first line) and the moderately clean (second line) regimes, it remains at least qualitatively true generally. Note, also, that utilizing results in our ideal dirty limit $(\gamma \ll 1 \ll \gamma_s)$ requires a strongly isotropic superconductor; otherwise, the suppression of $T_c$ is not captured correctly by our small pair-breaking parameter.

Reference~\cite{Kubo22} reports results on superconducting residual resistance, inversely proportional to the quality factor, within the Eilenberger formalism. To compare it with our approach based on DS model, we express the ideal clean $(\gamma\ll 1$, $\gamma_s \ll 1)$ and ideal dirty limit. Using the superconducting gap $\Delta_{00}$ of the clean BCS system (measured, e.g. by STM for $\Gamma=0$), we are left with
\begin{widetext}
\begin{equation*}
    \frac{Q_s(0)}{Q_{00}}=
    \begin{cases}
        \Big(\frac{\Delta_{00}}{\omega}\Big)^\frac{3}{2}\Big(\frac{2\Delta_{00}}{\Gamma}\Big)^2\Big[1-\frac{9}{2}\frac{\Gamma}{\Delta_{00}}-\left(\frac{3\pi}{8}-1\right)\frac{\Gamma_s}{\Delta_{00}}\Big] & \text{if}\quad \Gamma,\Gamma_s \ll \Delta_{00}\, \text{(ideal, clean)},\\
        \sqrt{\frac{2\Delta_{00}}{\Gamma_s}}\Big(\frac{\pi \Delta_{00}}{\omega}\Big)^\frac{3}{2}\Big(\frac{\Delta_{00}}{\Gamma}\Big)^2\Big[1-\left(\frac{3}{\pi}+\frac{7}{2}\right)\frac{\Gamma}{\Delta_{00}}-\frac{3}{\pi}\frac{\Delta_{00}}{\Gamma_s}\ln\left(\frac{\Gamma_s}{\Delta_{00}}\right)\Big] & \text{if}\quad \Gamma \ll \Delta_{00} \ll \Gamma_s\, \text{(ideal, dirty)},
    \end{cases}
\end{equation*}
\end{widetext}
where $Q_{00} = Q_{0}\sqrt{\Delta_0/\Delta_{00}}$. Note the $\ln(\Gamma_s/\Delta_{00})$ term, probably neglected in Ref.~\cite{Kubo22}.

\begin{figure}[h!]
    \includegraphics[width = 0.425\textwidth]{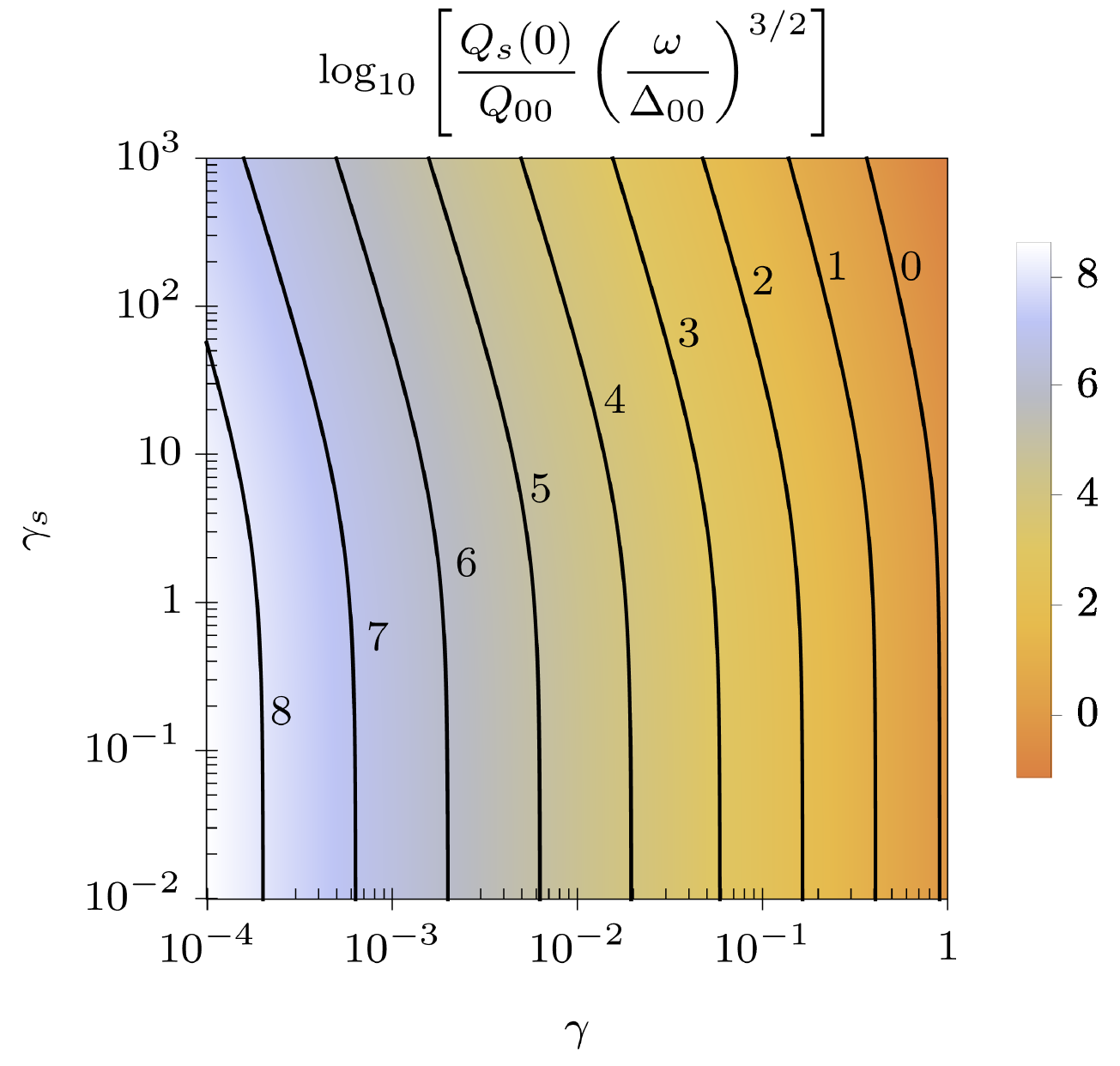}
    \caption{Quality factor at $T=\SI{0}{K}$ in the broad range of $\gamma$ and $\gamma_s$.
    }
    \label{fig:Qs0lQ00}
\end{figure}

Since the formulas above (otherwise useful in the specific regimes) may be cumbersome for the general understanding, we plot $Q_s(0)/Q_{00}$ defined by Eq.~\eqref{eq:QsT0} (together with Eq.~\eqref{eq:Re_sigma_T_0} and \eqref{eq:dens_frac}) as a function of scattering constants $\gamma$ and $\gamma_s$ in Fig.~\ref{fig:Qs0lQ00}. For simplicity, we multiply $Q_s(0)$ with the omnipresent factor of $\left(\omega/\Delta_{00}\right)^{3/2}$. The overall result confirms the different roles of pair-breaking and pair-conserving scattering. Considering the bent shape of the contours, it is easy to conclude that the dependence on $\gamma_s$ changes. Considering the fixed value of $\gamma \sim 10^{-3}$, at relatively small, but increasing $\gamma_s \lesssim 1$, the quality factor is continuously suppressed much less than in the case of increasing $\gamma_s \gtrsim 1$. This behavior highlights, among other things, the difference between the moderately clean and the ideal dirty regimes. As shown in Fig.~\ref{fig:Qs0lQ00}, this difference between small and large $\gamma_s$ scaling is present generally, considering relevant values of $\gamma\lesssim 1$.

\section{Conclusions and Discussion}\label{sec:Concl&Discuss}

The huge versatility of (usually niobium-based) SRF cavities implies the demand for a simple analysis of resonant frequency shift and quality. We have addressed this problem within the DS theory framework.

Focusing on the frequency shift shows that, despite omitting the inhomogeneity or anisotropy of the superconducting order parameter-related effects, our description provides qualitatively reasonable results with the same numerical complexity as the Mattis-Bardeen theory. Among other things, we have systematically identified various frequency shift regimes (Foot, Bump, Dip~\&~Bump and Dip) assuming low disorder in Sec.~\ref{subsec:imprints}. In addition, addressing probably the most relevant, moderately clean regime, we have described the frequency shift slope $\delta f'(\Theta)\lvert_{\Theta = 0}$ with the simple function of pair-conserving and pair-breaking scattering rates, analyzed in Fig.~\ref{Fig:Slope_mod_clean}. An analytical description of the width and depth of the resulting dip in the moderately clean regime clarifies the role of the pair-conserving and pair-breaking disorder, combined with the finite frequency range. It also allows for the natural interpretation utilizing relevant penetration depths on both sides of the superconductor-normal metal phase transition.

We have also discussed our findings concerning recent experiments and more complex theory approaches in Sec.~\ref{subsec:imprints}. To test the limits of the DS framework, we have compared and fit $\delta f(T)$ with the recently analyzed clean N-doped Nb sample in Fig.~\ref{Fig:comparison_and_fit}, and we also interpret the results. Considering the complexity of the phenomena together with the relevant temperature scale and simplicity of our approach, we have achieved compelling agreement with the experiment.

In Sec.~\ref{sec:Q} we have analyzed the quality factor mainly in the regime of low pair-breaking and low-temperature range. We discuss the nature of the present plateaus and their relevant temperature scale $T_p$. Considering $T=\SI{0}{K}$, we list simple analytical results describing $Q_s(0)$ in various regimes, discussing their most important properties. At a moderately clean regime, these results can be used as a proxy of $Q_s(T\lesssim T_p)$ at the range of discussed plateaus. 

Summarizing achieved results, we believe that our detailed analysis may serve as a convenient tool. In particular, if we think about fundamental particle acceleration, detection, and also preservation, considering quantum technologies.

\section*{Acknowledgments}

This work was supported by the Slovak Research and Development Agency under Contract no. APVV-23-0515 and by the European Union's Horizon 2020 research and innovation program under Marie Sk{\l}odowska-Curie Grant Agreement No.~945478.

\clearpage
\appendix

\section{Dip within two-fluid model}\label{app:Dip_2FM}

If we assume the two-fluid superconducting model \cite{Tinkham04}, assuming the conductivity at finite frequencies in the superconducting state is
\begin{equation*}
    \frac{\sigma_s}{\sigma_0} = \frac{1}{1-i\omega\tau}\left(\frac{T}{T_\mathrm{c}}\right)^4 + \frac{i}{\omega\tau}\left[1 - \left(\frac{T}{T_\mathrm{c}}\right)^4\right],
\end{equation*}
where the temperature dependence models the temperature evolution of concentrations of the normal and superconducting electrons below $T_c$. Above $T_c$ we consider just the normal electron contribution to the conductivity, resulting in Drude form, formulated by Eq.~\eqref{eq:Drude}. Since we are focused on the temperatures near $T_c$, it is convenient to define $\Theta\equiv 1-T/T_c \ll1$. In such a case the superconducting conductivity can be written as
\begin{align*}
    \frac{\sigma_s}{\sigma_0} &= \frac{1-4\Theta}{1-i\omega\tau}+i\frac{4\Theta}{\omega\tau},\\
    &\approx 1+i\frac{4\Theta}{\omega\tau} = \sqrt{1+\left(\frac{4\Theta}{\omega\tau}\right)^2}e^{i\arctan{\left(\frac{4\Theta}{\omega\tau}\right)}},
\end{align*}
where, in the second line, we assumed $\omega\tau\ll 1$, according to our considerations. Under the same circumstances $\sigma_n/\sigma_0 \approx 1$ and $\alpha_n=\pi/4$ in Eq.~\eqref{eq:Zn}. Expressing the frequency shift \eqref{eq:ResFreqShift}, we are left with
\begin{multline}\label{eq:f-shift-approx-precise}
    \frac{\delta f(\Theta)}{\Tilde{f}} = 1-\frac{\sqrt{2}\sin\left(\pi/4+\arctan\left(\frac{4\Theta}{\omega\tau}\right)/2\right)}{\left[1 + \left(\frac{4\Theta}{\omega\tau}\right)^2\right]^{\frac{1}{4}}}.
\end{multline}
Concentrating on the simplest case, when $\Theta\ll \omega\tau$, leads after linearization to $\delta f(\Theta)/\Tilde{f} \approx -2\Theta/(\omega\tau)$, which immediately reveals the slope of the dip close to $T_c$
\begin{equation*}
   \frac{\partial}{\partial \Theta} \left.\frac{\delta f(T)}{\Tilde{f}}\right\rvert_{\Theta = 0} \approx -\frac{2}{\omega\tau},
\end{equation*}
and its scaling on $\omega\tau$. This result implies that the dip is caused by the disorder's effects, and in principle allows the order of magnitude estimation of the scattering time~$\tau$ in relatively clean superconducting cavities. 

Next, focusing on the other root of $\delta f(\Theta_0)$ = 0 in Eq.~\eqref{eq:f-shift-approx-precise}, resulting in the value of the dip width estimation $\Theta_0\approx \omega\tau\pi/8$, representing another way of estimating the scattering time. For completeness, let us complete our rough analysis with the constant dip depth approximation $\delta f(\Theta_\mathrm{dip}=\Theta_0/2)\approx -0.13 \tilde{f}$. We find it interesting that our oversimplified analysis shows the following. The estimate of $f(\Theta_\mathrm{dip})$ roughly agrees with our previous findings within the moderately clean regime, based on a more sophisticated DS theory approach presented in Sec.~\ref{subsec:imprints}. Considering reasonable numbers for $\tilde{f}$ (therefore, $X_n$, $G$, and $f$), this estimate is also in at least qualitative agreement with the experimentally observed $\SI{}{\kilo\hertz}$ scale (in absolute units). However, within the two-fluid model, the depth of the dip in scale $\tilde{f}$ is insensitive to the disorder, since the right-hand side of Eq.~\eqref{eq:f-shift-approx-precise} depends on the ratio of $4\Theta_\mathrm{dip}/(\omega\tau) = \pi/4$.

\begin{figure}[h!]
    \centering
    \includegraphics[width=0.4\textwidth]{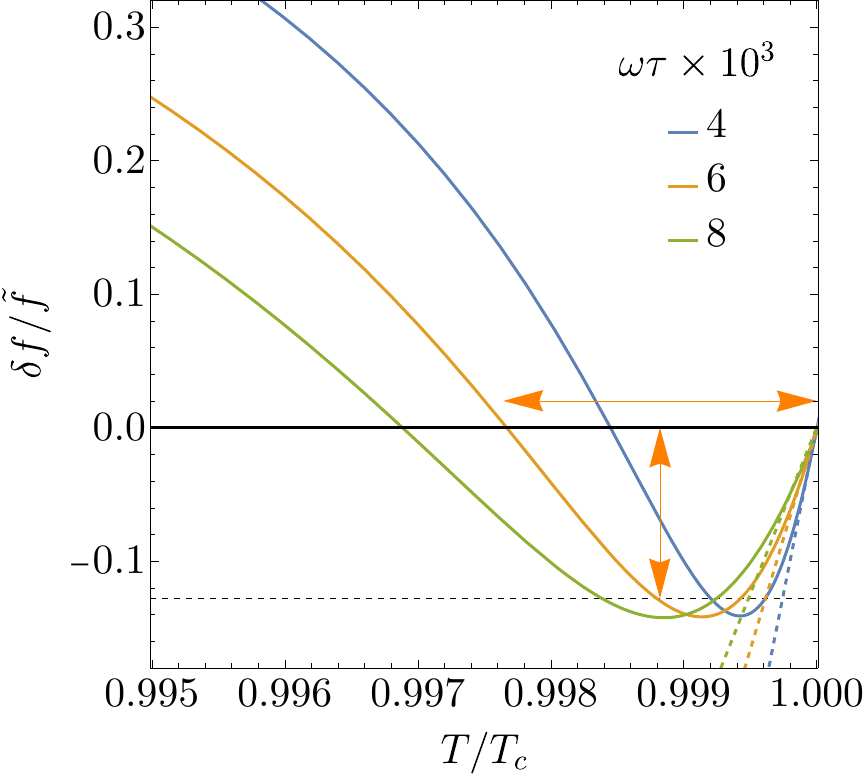}
    \caption{Graph showing dip in frequency shift near the critical temperature $T_\mathrm{c}$ for different frequencies. Arrows show the estimated width and depth for $\omega\tau = 6\times\num{e-3}$(orange curve). We may note that depth is approximately the same for all values, as shown by the estimate corresponding to the dotted line.}
    \label{fig:dip}
\end{figure}

In Fig.~\ref{fig:dip}, we show a graphic interpretation of our very simple model according to Eq.~\ref{eq:f-shift-approx-precise} showing {\it i)} a slope close to $T_c$ and {\it ii)} dip width and dip depth estimate assuming a couple of values of $\omega\tau$.

Previous results may also be formulated in terms of the London penetration depth $\lambda_L(T)$ from Eq.~\eqref{eq:Crit} and the depth of the skin effect in the normal state $\delta^2 = 2\lambda_{L0}^2/(\omega\tau)$ utilizing\footnote{Note that $\delta \gg \lambda_{L0}$, since $\omega\tau \ll 1$.} the ratio
\begin{equation*}
    \left(\frac{\delta}{\lambda_L(T)}\right)^2 =
    \frac{2}{\omega\tau}\frac{n_s(T)}{n}.
\end{equation*}
The superfluid fraction (and also $\Theta$ in the vicinity of $T_c$) can be expressed in terms of $\delta/\lambda_L$. We can note that $\delta f$ described in terms of Eq.~\eqref{eq:f-shift-approx-precise} depends only on the ratio $4\Theta/\omega\tau = (\delta/\lambda_L)^2/2$, and therefore can be described purely by the comparison of different penetration depths. For the linearized frequency shift and $\lambda_L(\Theta_0)$ we are left with
\begin{equation*}
    \frac{\delta f}{\tilde{f}}\approx-\left(\frac{\delta}{2\lambda_L}\right)^2=-\left(\frac{\delta}{\lambda_{L0}}\right)^2\Theta,\quad
    {\lambda_L(\Theta_0)} = \frac{\delta}{\sqrt{\pi}}.
\end{equation*}
The ratio of penetration depths reveals the natural scale of the large negative slope of the frequency shift at $T_c$. In addition, the natural scale of the dip width is indicated as $\delta f=0$, when $\lambda_L$ is comparable with $\delta$. For that to happen, the superfluid fraction must be very small ($n_s/n_0 \approx \omega\tau$), requiring temperatures to be very close to $T_c$. Realizing $\Theta_{0}=\pi \Theta_{eq.}$, where $\Theta_{eq.} = \omega\tau/8$ marks the temperature where $\lambda_L=\delta$, we see the dip results from the comparable depth length scales in superconducting and normal states.

\section{Superfluid fraction at $T=\SI{0}{K}$}\label{Appendix:App_sc_frac}
The general form of the superfluid fraction utilized in the Sec.~\ref{subsec:imprints} from Ref.~\cite{Herman17b} is
\begin{widetext}
\begin{equation}\label{eq:dens_frac}
    n_s(0)/n = \begin{cases}
        \frac{1}{\gamma_s}\bigg[
        \arctan(1/\gamma)-\frac{1}{\sqrt{1-\gamma_s^2}}\left(\arccos \gamma_s + \arctan\frac{\sqrt{1-\gamma_s^2}}{\gamma} - \arctan\frac{\sqrt{1-\gamma_s^2}\sqrt{1+\gamma^2}}{\gamma \gamma_s}\right)\bigg],\, \mathrm{if} \, \gamma_s<1\\
        \frac{1}{\gamma_s}\bigg[
        \arctan(1/\gamma)-\frac{1}{\sqrt{\gamma_s^2-1}}\ln\frac{\left(\gamma_s+\sqrt{\gamma_s^2-1}\right)\left(\gamma+\sqrt{\gamma_s^2-1}\right)}{\gamma\gamma_s+\sqrt{\gamma_s^2-1}\sqrt{\gamma^2+1}}\bigg],\, \mathrm{if}\, \gamma_s \geq 1.
    \end{cases}
\end{equation}
\end{widetext}

Assuming $\Gamma_s < \Delta_0$ and expanding to the linear order of $\Gamma$, we are left with the known equation \cite{Herman21}
\begin{equation}\label{eq:dens_frac_1}
    \frac{n_s(0)}{n} = \frac{1}{\gamma_s}\bigg[\frac{\pi}{2} - \frac{\arccos{\gamma_s}}{\sqrt{1 - \gamma_s^2}}\bigg] - \frac{\gamma}{\gamma_s + 1}.
\end{equation}

\section{Dip in ideal dirty limit}\label{Appendix:ideal dirty}
{(\it Ideal dirty limit)} In this appendix, we follow up on the analysis of $\delta f(T)$ from Sec.~\ref{Sec:DipAnalysis}. We focus on the regime $\hbar\omega\ll\Gamma\ll\Delta(0) \approx \Delta_{00}\ll\Gamma_s$. Similar to the moderately clean regime described in Sec.~\ref{Sec:DipAnalysis}, we utilize equations from Ref.~\cite{Herman21} close to $T_c$ in order of $\Delta(T)^2$ in the dirty limit, specifically
\begin{align*}
    \frac{\sigma_s'(T)}{\sigma_0} &\approx 1+\frac{\pi}{8}\frac{\Delta(T)^2}{\Gamma T_c}, \\ \frac{\sigma_s''(T)}{\sigma_0} &\approx \frac{\pi\Delta(T)}{\omega}\tanh\left(\frac{\Delta(T)}{2T}\right) \approx\frac{\pi}{\omega}\frac{\Delta(T)^2}{2T}.
\end{align*}
Strictly speaking, the relation for $\sigma''_s(T)$ is true in the dirty limit with $\Gamma=0$; however, the effect of small, but finite $\Gamma$ shows only a very small deviation from our form~\cite{Herman21}. Considering our parameter regime, we are left with
\begin{align}
    \frac{\sigma_s(T)}{\sigma_0} &\approx 1 + \frac{\pi}{2}\bigg(i+\frac{\omega}{4\Gamma}\bigg)\frac{\Delta^2(T)}{\omega T_c},\nonumber \\
    &\approx 1 + \gamma_{\sigma}\frac{\Delta_{00}}{\omega}\bigg(i+\frac{\omega}{4\Gamma}\bigg)\bigg(1-\frac{T}{T_{c0}}\bigg)\label{eq:sigma_s_drt_lmt_cls_Tc},
\end{align}
where, in the second line, we used the well-known BCS limit for $\Delta(T)$ \cite{Parks75}. From this result, it is easy to identify the slope of $\delta f(T)$ in the absolute vicinity of the $T\approx T_c\approx T_{c0}$ throughout the first term of the Taylor series in variable $\left(1-T/T_{c0}\right)$ as
\begin{equation}\label{eq:slope}
    \frac{\delta f(T)}{\tilde f} \approx \frac{\gamma_{\sigma}}{2}\frac{\Delta_{00}}{\omega}\bigg(\frac{\omega}{4\Gamma}-1\bigg)\bigg(1-\frac{T}{T_{c0}}\bigg).
\end{equation}
Note the role of $\omega/(4\Gamma)\ll1$ (from assumption), which makes the slope positive in $T$ (implying dip), and increasing this ratio flattens the dip close to $T_c$.

A very rough estimate of the dip's width can be obtained in the very same way as for the moderately clean regime, as a solution of $\delta f (T_w) = 0$ for $T_w \neq T_{c0}$, by using Eq.~\eqref{eq:sigma_s_drt_lmt_cls_Tc} in the local form of Eq.~\eqref{eq:delta_f_loc} assuming $\omega\tau \rightarrow 0$ in normal state related properties. Considering the series up to the second order in $\Theta$ reveals
\begin{equation*}
    \Theta_{\mathrm{dip}} \approx \frac{\omega}{\gamma_{\sigma}\Delta_{00}}\bigg(\frac{2}{3}-\frac{\omega}{2\Gamma}\bigg),\quad \frac{\delta f\left(T_\mathrm{dip}\right)}{\tilde{f}}\approx -\frac{1}{6}\left(1-\frac{\omega}{\Gamma}\right),
\end{equation*}
holding true for $\omega\ll \Gamma$. This result reveals the characteristic temperature and depth of the dip. It suggests a decrease in the absolute value of the depth, as well as a narrowing of the dip width, with the increasing ratio of $\omega/\Gamma$.
\begin{figure}[h!]
    \includegraphics[width = 0.45\textwidth]{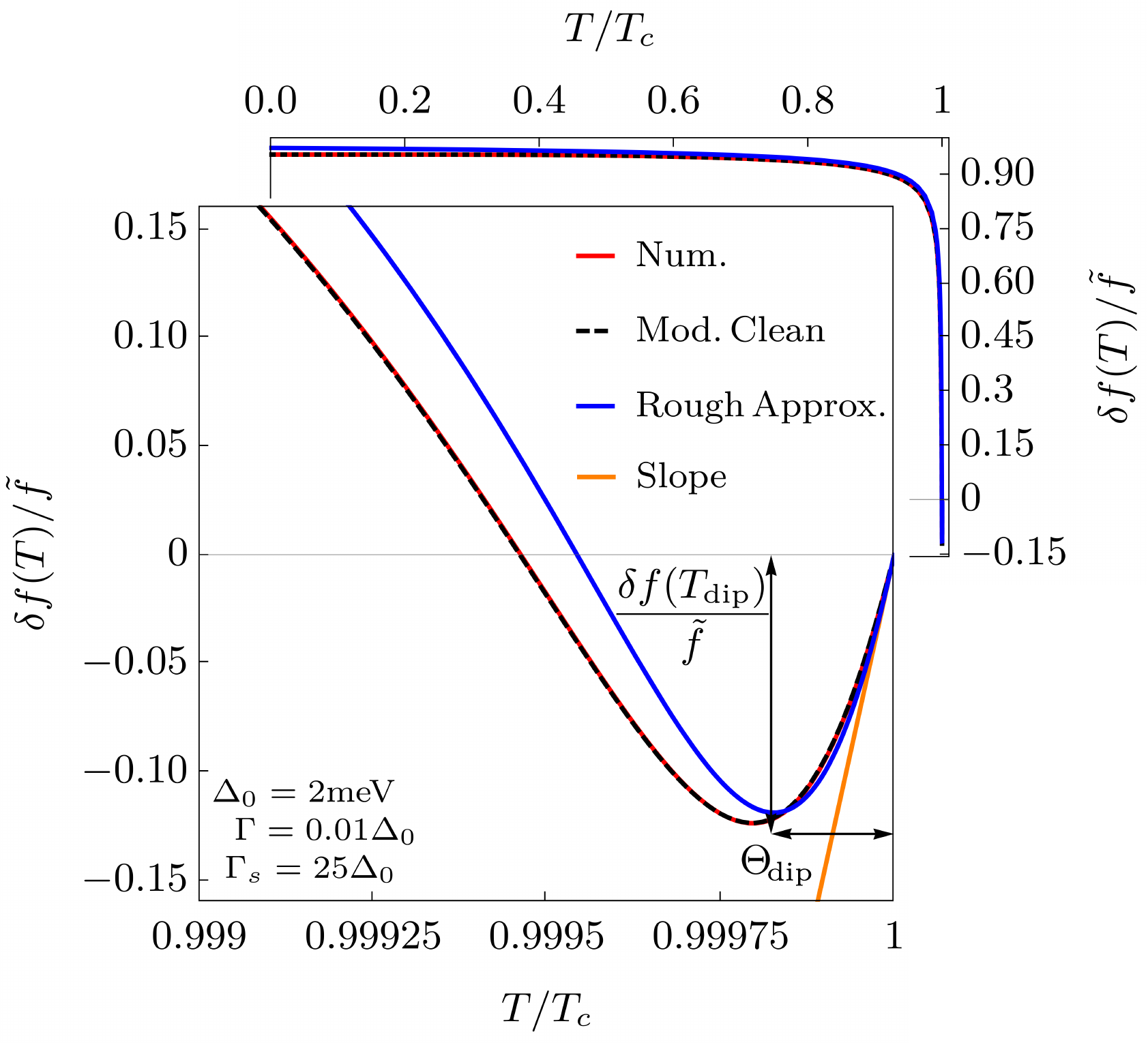}
    \caption{Dip in the dirty limit of DS model.}
    \label{fig:DirtyLimit_Analytics}
\end{figure}

In Fig.~\ref{fig:DirtyLimit_Analytics} we show the rough estimates for $\Theta_{\mathrm{dip}}$ and depth by black arrows. The red curve corresponds to complete numerics by combining Eq.~\eqref{eq:ResFreqShift} with Eq.~\eqref{eq:Sigma_Num}. For control purposes, the black dashed curve corresponds to Eq.~\eqref{eq:ResFreqShift} using $\sigma_s(\omega)$ in the form of the moderately clean superconductors \cite{Herman21}, which can also be utilized in this scenario. The orange curve represents the slope from Eq.~\eqref{eq:slope}. The blue curve represents the rough approximation that combines Eq.~\eqref{eq:ResFreqShift} and Eq.~\eqref{eq:sigma_s_drt_lmt_cls_Tc}.

To summarize, in the considered ideal dirty limit, decreasing pair-breaking scattering leads to an overall smaller dip effect in the resonant frequency shift. Last but not least, since we are in the dirty limit, the response in the normal state can be considered in the local limit since $|q_0|\ell$ decreases due to decreasing (increasing) $\ell$ ($\delta$). In addition, from the superconducting side, the penetration depth increases \cite{Herman18}, meanwhile the size of the Cooper pair decreases \cite{Herman23} with the (assumed dominant) $\Gamma_s$. Therefore, we do not have to worry about the dip being extremely close to $T_c$.

\section{Dip comparison for other resonant frequencies}\label{Appendix:Dip_Other_freq}

\begin{figure}[h!]
    \centering
    \includegraphics[width=0.32\textwidth]{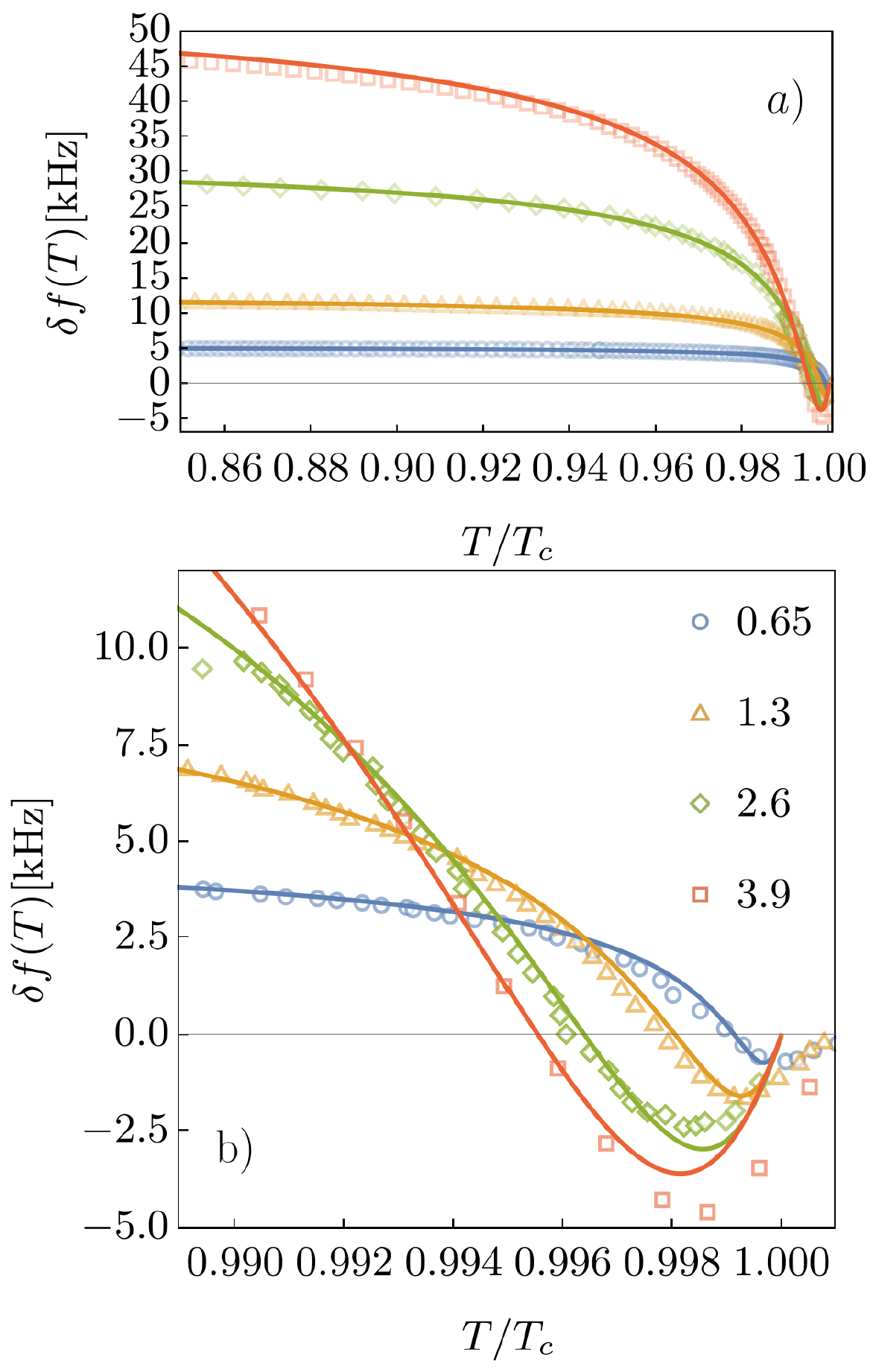}
    \caption{Resonant frequencies are presented in units of GHz in the legend, discrete symbols correspond to digitalized data, and the curves to the theory calculations. Bottom-figure b) focuses on the detail close to the critical temperature $T_c$.}
    \label{fig:Res_4Freq}
\end{figure}

To complete our analysis shown in Fig.~\ref{Fig:comparison_and_fit}, we show the fitting result for all $4$ cavities with different resonant frequencies from Ref.~\cite{Ueki22} in Fig.~\ref{fig:Res_4Freq}. The results fitted on the interval of $0.96\leq T/T_c\leq1$ were obtained by optimizing values of $T_c$ and $R_n$ (determining $\Gamma$ and $\Gamma_n$). Alternatively to fitting (however leading to slightly less-accurate results) can be done by estimating individual values of $T_c$ and taking the experimental value of $X_n = R_n$ (assuming local limit).

All interesting parameters resulting from fitting are given in Table~\ref{tab:Res_4Freq}.

\begin{center}
\begin{table}[h!]
\begin{tabular}{ |c|c|c|c|c|c| }
 \hline
 $f[\SI{}{GHz}]$ & $R_n[\SI{}{m\Omega}]$ & $T_c[\SI{}{K}]$ & $\Gamma[\SI{}{meV}]$ & $\Gamma_n[\SI{}{meV}]$ & $\Delta_0[\SI{}{meV}]$\\
 \hline
 0.65 & 4.451 & 8.959 & 0.044 & 1.330 & 1.546\\
 \hline
 1.3 & 5.642 & 8.875 & 0.054 & 1.068 & 1.536\\
 \hline
 2.6 & 7.506 & 9.061 & 0.032 & 0.945 & 1.558\\
 \hline
 3.9 & 9.377 & 9.135 & 0.024 & 0.983 & 1.567\\
 \hline
\end{tabular}
    \caption{Experimental values of $f$, determined $R_n$, $T_c$ and corresponding parameters $\Gamma$, $\Gamma_n$ and $\Delta_0$.}
    \label{tab:Res_4Freq}
\end{table}
\end{center}

The overall message is clear. The basic characteristics of the frequency shift are visible in the theory and copy those seen in the experiment. The resulting parameters are consistent with the moderately clean regime for all the cavities, as the overall scattering rate is comparable to the size of the zero-temperature gap. The fitted values of $T_c$ are in agreement (two in qualitative, two in perfect) with the experimental findings considered in Tab.~I of Ref.~\cite{Ueki22}, and the determined values of $R_n$ differ by less than $8\%$ from the experimental values in the same table. 

Of course, it is probably a bit naïve to expect the fit to be absolutely perfect. First, experimentally, it is perhaps hard to exclude the effect of the temperature gradient on the surface of the cavity. Second, however, even if, mainly, we will get to the limits of our model based on the CPA (mean-field) approach, where the characteristic behavior does not have to match the inhomogeneity profile of the cavity material perfectly. Still, even if we consider the imperfections of the used approximation(s), we see that the utilized approach can determine the basic SRF cavity characteristics with comparable accuracy to the more sophisticated approaches (comparing Fig.~\ref{fig:Res_4Freq} with Fig.~4 in Ref.~\cite{Ueki22} or Fig.~3 in Ref.~\cite{Zarea23a}).

The data in our Fig.~\ref{Fig:comparison_and_fit} originate from Ref.~\cite{Zarea23a}, where the dip was plotted only for the case of $2.6$ GHz. We have preferably chosen this source for the main text analysis due to the best detail of the dip, most suitable for digitalization, and also for the clear information about the corresponding $T_c$ and the value of the superconducting gap $\Delta_0$ at $T=\SI{0}{K}$.

\section{Ballpark estimate of $R_s(0)$ in case of Nb$_3$Sn}\label{Appendix:App_Nb3Sn}

In the following, we apply our approach to very roughly estimate the pair-breaking contribution to the surface resistance $R_s$ at $T=\SI{0}{K}$ for the promising SRF material Nb$_3$Sn. As in the case of Nb considered and analyzed in the main text, even in this case, we will, for simplicity, take the anisotropic properties of the order parameter only throughout the presence of nonzero pair-breaking scattering. We also neglect that the analyzed Nb$_3$Sn is (considering the atomic Sn content) in the strong coupling regime \cite{Godeke06} and consider the material as a standard low-temperature conventional superconductor. Our approximation is based on the Dynes-like character of the Nb$_3$Sn point contact tunneling conductance measurements \cite{PosenPHDThesis}, revealing a relatively sharp gap at low temperatures, suggesting a small influence of the pair-breaking effect.

To specify, we will utilize properties of Nb$_3$Sn cavity 2 from test 6 reported in \cite{PosenPHDThesis} and Tab.~7.2 therein.
We consider the resonant frequency $f=~\SI{1.3}{GHz}$, clean limit zero Kelvin temperature London penetration depth $\lambda_{L0}=\SI{89}{nm}$, coherence length (roughly equal to the size of the Cooper pair) $\xi_0 = \SI{7}{nm}$, and mean free path $\ell = \SI{4.8}{nm}$. The routine check reveals that already $\lambda_{L0}\gg \ell$, justifying the local response to the external electromagnetic field.

First, we can immediately estimate the overall scattering rate $\gamma_n =\sqrt{2} \xi_0/\ell\approx2$, showing the intermediate disorder regime. This may be surprising, since naïvely, judging only by the lower $\ell$ than in Nb samples, one could expect a dirty limit. However, slightly smaller $v_F\sim\SI{0.1}{Mm/s}$ and larger $\Delta_0$, causing smaller $\xi_0$ point toward the mentioned intermediate disorder regime.

Next, for the rough estimation purposes, we naïvely identify the reduction of the atomic Sn content (reducing the $T_c$ and the value of the superconducting gap \cite{PosenPHDThesis, Posen14} in Nb$_3$Sn) as a plausible pair-breaking tuning knob in the linear regime (when $\delta T_c\ll T_{c0}$). In this sense, we take the value of the maximum possible critical temperature $T_{c,\, max.} = \SI{18.3}{K}$ \cite{PosenPHDThesis, Hanak64} as the critical temperature of the clean system $T_{c0}$. For the actual critical temperature of the system, we take $T_c = \SI{17.9}{K}$, being on the lower border of the reported accuracy interval $18\pm 0.1\SI{}{K}$ \cite{PosenPHDThesis}, allowing for the maximal effect of the pair-breaking within our considerations. Using these values within Eq.~\eqref{eq:gamma_lin} leads to $\gamma\approx 2 \times10^{-2}$, which together with the already realized value of $\gamma_s$ nicely complements the intermediate disorder regime with a naturally small effect of the pair-breaking scattering. Alternatively, $\gamma$ could be determined in a more sophisticated way, by analyzing the point contact tunneling measurements, reported in Refs.~\cite{PosenPHDThesis, Posen14}. One way or another, our estimated value of $\gamma$ qualitatively fits the presence of relatively sharp gap edges at low temperatures. The considered value of the $T_c$ reveals the gap \cite{PosenPHDThesis} $\Delta_0 = 2.25 k_BT_c \approx \SI{3.5}{meV}$.

Next, considering $R_s = R_n Q_n/Q_s$ in $T\rightarrow\SI{0}{K}$ limit, our realized regime of $\gamma\ll1$ and $\gamma_n\sim1$ together with equations Eqs.~\eqref{eq:Rn_or_X_n},~\eqref{eq:QsT0},~\eqref{eq:Re_sigma_T_0}, 
and~\eqref{eq:dens_frac} reveals
\begin{equation}\label{eq:Rs(0)ex}
    R_s(0) = \frac{\mu_0 \hbar \lambda_{L0}(\omega\gamma)^2}{4\Delta_0(1+\gamma_s)}\left(\frac{n}{n_s(0)}\right)^{3/2}\approx \SI{0.2}{n\Omega}.
\end{equation}
\vspace{0.2cm}

To compare, the residual resistance reported in Ref.~\cite{PosenPHDThesis} is $R_{res}\approx \SI{8.5}{n\Omega}$, being one order of magnitude higher than our result. The reasons for this discrepancy can certainly be identified in the oversimplification of the Nb$_3$Sn superconducting properties description by completely ignoring the anisotropy of the order parameter. Another reason is probably related to the fact that in our approach, the only contributing factor to $R_s(0)$ is coming from the pair-breaking scattering. At the same time, in the experiment, there can be other contributing factors such as trapped vortices \cite{Gurevich17}, two-level-system defects \cite{Muller19}, or nonequilibrium quasiparticles \cite{Visser14}.

However, despite the mentioned difference, we do not find the ballpark estimate absolutely off or inconsistent, especially if we take into account the various involved energy scales of $\Delta_0 \sim\Gamma_n\sim \SI{}{meV}$, $\Gamma\sim \SI{10}{\micro eV}$ and $\hbar\omega\sim\SI{}{\micro eV}$, and that the numerical difficulty of the ballpark estimate requires {\it i)} basic and well-known characteristics of the material, {\it ii)} our main text resulting equations and  (at worst) {\it iii)} an average calculator.

As a final remark, note that the $(1+\gamma_s)$ and the superfluid fraction $n_s(0)/n$ related factors in Eq.~\eqref{eq:Rs(0)ex} are of order of magnitude $\sim1$ and do not change the order of magnitude estimate in the considered regime of $\Gamma\ll \Delta_0\lesssim \Gamma_n$. In case of $\gamma\ll1$ and $\gamma_s\ll1$ we are left with even simpler result of 
\begin{equation*}
     R_s(0) = \frac{\mu_0 \hbar \lambda_{L0}}{\Delta_0}\left(\frac{\omega\gamma}{2}\right)^2,
\end{equation*}
showing dependence purely on the resonant frequency, pair-breaking scattering rate, and fundamental superconducting properties.

\end{document}